\documentclass[12pt]{article}
\usepackage{a4wide}
\usepackage{latexsym}
\usepackage{cite}
\usepackage{graphicx}

\usepackage{pslatex}
\usepackage[latin1]{inputenc}
\usepackage[T1]{fontenc}

\def\bq{\begin{eqnarray}}
\def\eq{\end{eqnarray}}
\def\l{\langle}
\def\r{\rangle}

\begin{document}

\thispagestyle{empty}

\begin{flushright}
  MZ-TH/07-14 \\
%  version of \today \\
\end{flushright}

\vspace{1.5cm}

\begin{center}
  {\Large\bf 
   Parton showers from the dipole formalism\\
  }
  \vspace{1cm}
  {\large 
    Michael Dinsdale, Marko Ternick and Stefan Weinzierl\\
  \vspace{1cm}
      {\small \em Institut f{\"u}r Physik, Universit{\"a}t Mainz,}\\
      {\small \em D - 55099 Mainz, Germany}\\
  } 
\end{center}

\vspace{2cm}

% abstract ---------------------------------------
\begin{abstract}\noindent
  {
We present an implementation of a parton shower algorithm 
for hadron colliders and electron-positron colliders
based on the dipole factorisation formul{\ae}.
The algorithm treats initial-state partons on equal footing with final-state partons.
We implemented the algorithm for massless and massive partons.
   }
\end{abstract}

\vspace*{\fill}

% main text ------------------------------------
\newpage

\section{Introduction}

Event-generators like Pythia\cite{Sjostrand:2006za,Bertini:2000uh}, 
Herwig\cite{Corcella:2000bw,Gieseke:2003hm} 
or Sherpa\cite{Gleisberg:2003xi}
are a standard tool in high energy particle physics.
In these tools the physics of particle collisions is modelled by a simulation with different
stages -- hard scattering, parton showering, hadronisation -- to name the most important ones.
The hard scattering process is calculable in perturbation theory.
The same holds -- in theory at least -- for the parton showering process, the relevant scales 
are still large enough for perturbation theory to be applicable.
In practise however, one is forced into approximations due to the large parton multiplicities.
These approximations are derived from the behaviour of the matrix elements in singular regions.
The matrix elements become singular in phase space regions corresponding to the emission
of collinear or soft particles.
The first showering algorithms started from the collinear factorisation of the matrix elements 
and approximated colour interference effects through angular ordering\cite{Marchesini:1983bm,Webber:1983if}.
An exception is the algorithm implemented in 
Ariadne\cite{Gustafson:1986db,Gustafson:1987rq,Andersson:1989ki,Andersson:1988gp,Lonnblad:1992tz},
which is based on a dipole cascade picture.
Most shower algorithms are in the collinear limit accurate to the leading-logarithmic approximation.
Extensions to the next-to-leading logarithmic approximation have been 
studied in\cite{Kato:1986sg,Kato:1988ii,Kato:1990as,Tanaka:2005rm}.

Recent years have witnessed significant developments related to shower algorithms,
including procedures to match parton showers to fixed-order tree-level 
matrix elements\cite{Catani:2001cc,Krauss:2002up,Schalicke:2005nv,Mangano:2001xp,Mrenna:2003if}
and methods to combine parton showers with next-to-leading order 
matrix elements\cite{Baer:1991qf,Baer:1992ca,Friberg:1999fh,Mrenna:1999mq,Potter:2000an,Potter:2001ej,Dobbs:2001gb,Dobbs:2001dq,Collins:2001fm,Frixione:2002ik,Kramer:2003jk,Soper:2003ya,Nason:2004rx,Nagy:2005aa,Nagy:2006kb,Nagy:2007ty,Kramer:2005hw,Kurihara:2002ne,Odaka:2007gu,Giele:2007di,Frixione:2007nu,Frixione:2007nw,Latunde-Dada:2006gx,LatundeDada:2007jg}.
The shower algorithms in Pythia, Herwig and Ariadne have been 
improved\cite{Sjostrand:2004ef,Gieseke:2003rz,Lonnblad:2001iq}
and new programs like the shower module
Apacic++\cite{Krauss:2001iv,Krauss:2005re}
of Sherpa have become available.
Other improvements include the study of uncertainties in parton showers\cite{Gieseke:2004tc,Stephens:2007ah,Bauer:2007ad},
as well as showers in the context of the soft-collinear effective theory \cite{Bauer:2006mk}.

Of particular importance is the matching of parton showers with next-to-leading order
matrix elements.
The pioneering project MC@NLO\cite{Frixione:2002ik,Frixione:2003ei,Frixione:2005vw,Frixione:2006gn,Frixione:2007zp} 
used an existing shower program (Herwig)
and adapted the NLO calculation to the shower algorithm, at the expense of sacrificing the correctness 
in certain soft limits.
It is clear that a better but more labour-intensive approach 
would adapt the shower algorithm to NLO calculations.
Nowadays in NLO computations the 
dipole subtraction method\cite{Catani:1997vz,Catani:1997vzerr,Dittmaier:1999mb,Phaf:2001gc,Catani:2002hc}
is widely used.
Nagy and Soper\cite{Nagy:2005aa,Nagy:2006kb}
proposed to build a shower algorithm from the dipole subtraction terms.

In this paper we report on an implementation of a shower algorithm based 
on the dipole formalism as suggested by Nagy and Soper. 
We take the dipole splitting functions as the splitting functions which generate the parton shower.
In the dipole formalism, a dipole consists of an emitter-spectator pair, which emits a third particle, soft or 
collinear to the emitter.
The formalism treats initial- and final-state partons on the same footing. 
In contrast to other shower algorithms, no distinction is made between final- and initial-state
showers.
The only difference between initial- and final-state particles occurs in the kinematics.
In the implementation we have the four cases
final-final, final-initial, initial-final and initial-initial corresponding to the possibilities of the particles
of the emitter-spectator-pair to be in the initial- or final-state.
All four cases are included, therefore the shower can be used 
for hadron colliders and electron-positron colliders.
We implemented the shower for massless and massive partons.
Initial-state partons are however always assumed to be massless.
We use spin-averaged dipole splitting functions.
The shower algorithm is correct in the leading-colour approximation.
As the evolution variable we use the transverse momentum in the massless case, and a variable
suggested in \cite{Dokshitzer:1995qm,Gieseke:2003rz} for the massive case.
The variable for the massive case reduces to the transverse momentum in the massless limit.
Schumann and Krauss report on a similar but separate implementation of a parton shower algorithm based on the dipole
formalism \cite{Schumann}.

This paper is organised as follows:
In section~\ref{sect:notation} we review basic facts about the colour decomposition of QCD amplitudes
and the dipole formalism.
In section~\ref{sect:shower} we discuss the shower algorithm.
In section~\ref{sect:results} we present numerical results from the parton shower simulation program.
Finally, section~\ref{sect:summary} contains the summary.
Technical details can be found in the appendix.
Appendix~\ref{appendix:final_final} discusses the case of a 
massless final-state emitter and a massless final-state spectator in detail.
Appendix~\ref{sect:insertion} describes the construction 
of the four-momenta of the $(n+1)$-particle state in all cases.
This appendix is also useful in the context of a phase-space generator for the real emission part
of NLO computations.

% ------------------------------------------------------------------

\section{QCD amplitudes and the dipole formalism}
\label{sect:notation}

In this section we briefly review the colour decomposition of QCD amplitudes
and the dipole formalism.

\subsection{Colour decomposition}
\label{sect:colour_decomp}

In this paper we use the normalisation
\bq
 \mbox{Tr}\;T^a T^b & = & \frac{1}{2} \delta^{a b}
\eq
for the colour matrices.
Amplitudes in QCD may be decomposed into group-theoretical factors (carrying the colour structures)
multiplied by kinematic functions called partial amplitudes
\cite{Cvitanovic:1980bu,Berends:1987cv,Mangano:1987xk,Kosower:1987ic,Bern:1990ux}. 
The partial amplitudes are gauge-invariant objects. 
In the pure gluonic case tree level amplitudes with $n$ external gluons may be written in 
the form
\bq
\label{colour_decomp_pure_gluon}
{\cal A}_{n}(1,2,...,n) & = & \left(\frac{g}{\sqrt{2}}\right)^{n-2} \sum\limits_{\sigma \in S_{n}/Z_{n}} 
 \delta_{i_{\sigma_1} j_{\sigma_2}} \delta_{i_{\sigma_2} j_{\sigma_3}} 
 ... \delta_{i_{\sigma_n} j_{\sigma_1}}  
 A_{n}\left( \sigma_1, ..., \sigma_n \right),
\eq
where the sum is over all non-cyclic permutations of the external gluon legs.
The quantities $A_n(\sigma_1,...,\sigma_n)$, called the partial amplitudes, contain the 
kinematic information.
They are colour-ordered, e.g. only diagrams with a particular cyclic ordering of the gluons contribute.
The choice of the basis for the colour structures is not unique, and several proposals
for bases can be found in the literature \cite{DelDuca:1999rs,Maltoni:2002mq}.
Here we use the ``colour-flow decomposition'' \cite{Maltoni:2002mq,Weinzierl:2005dd}.
This basis is obtained by replacing every contraction over an index in the adjoint
representation by two contractions over
indices $i$ and $j$ in the fundamental representation:
\bq
V^a E^a & = & V^a \delta^{ab} E^b = V^a \left( 2 T^a_{ij} T^b_{ji} \right) E^b
 =  \left( \sqrt{2} T^a_{ij} V^a \right) \left( \sqrt{2} T^b_{ji} E^b \right).
\eq
As a further example we give the
colour decomposition for a tree amplitude with a pair of quarks:
\bq
\label{colour_decomp_quark_gluon}
{\cal A}_{n+2}(q,1,2,...,n,\bar{q}) 
& = & \left(\frac{g}{\sqrt{2}}\right)^{n} 
 \sum\limits_{S_n} 
 \delta_{i_q j_{\sigma_1}} \delta_{i_{\sigma_1} j_{\sigma_2}} 
 ... \delta_{i_{\sigma_n} j_{\bar{q}}} 
A_{n+2}(q,\sigma_1,\sigma_2,...,\sigma_n,\bar{q}),
\eq
where the sum is over all permutations of the gluon legs. 
The tree amplitude with a pair of quarks, $n$ gluons and an additional lepton pair
has the same colour structure as in eq.~(\ref{colour_decomp_quark_gluon}).
In squaring these amplitudes a colour projector
\bq
 \delta_{\bar{i} i} \delta_{j \bar{j}} - \frac{1}{N_c} \delta_{\bar{i} \bar{j} } \delta_{j i}
\eq
has to applied to each gluon.
In these examples we have two basic colour structures, a colour cluster described by the ``closed string''
\bq
 \delta_{i_{\sigma_1} j_{\sigma_2}} \delta_{i_{\sigma_2} j_{\sigma_3}} 
 ... \delta_{i_{\sigma_n} j_{\sigma_1}}  
\eq
and a colour cluster corresponding to the ``open string''
\bq
 \delta_{i_q j_{\sigma_1}} \delta_{i_{\sigma_1} j_{\sigma_2}} 
 ... \delta_{i_{\sigma_n} j_{\bar{q}}}.
\eq
Born amplitudes with additional pairs of quarks have a decomposition in colour factors, which are products
of the two basic colour clusters above.
The colour factors in eq.~(\ref{colour_decomp_pure_gluon}) and eq.~(\ref{colour_decomp_quark_gluon}) 
are orthogonal to leading order in $1/N_c$.

\subsection{The dipole formalism}
\label{sect:dipole_formalism}

The starting point for the calculation of an observable $O$ in
hadron-hadron collisions in perturbation theory is the following formula:
\bq
\l O \r & = & \int dx_1 f(x_1) \int dx_2 f(x_2) \frac{1}{2 K(\hat{s})}
             \frac{1}{\left( 2 J_1+1 \right)}
             \frac{1}{\left( 2 J_2+1 \right)} 
             \frac{1}{n_1 n_2}
 \nonumber \\
 & &
             \int d\phi_{n}\left(p_1,p_2;p_3,...,p_{n+2}\right)
             O\left(p_1,...,p_{n+2}\right)
             \left| {\cal A}_{n+2} \right|^2.
\eq
This equation gives the contribution from the $n$-parton final state.
The two incoming particles are labelled $p_1$ and $p_2$, while $p_3$ to $p_{n+2}$ denote
the final state particles.
$f(x)$ gives the probability of finding a parton $a$ with momentum fraction $x$ inside
the parent hadron $h$.
A sum over all possible partons $a$ is understood implicitly.
$2K(s)$ is the flux factor,
$1/(2J_1+1)$ and $1/(2J_2+1)$ correspond
to an averaging over the initial helicities and
$n_1$ and $n_2$ are the number of colour degrees  of the initial state particles.
$d\phi_n$ is the phase space measure for $n$ final state particles, including (if appropriate) the identical particle factors.
The matrix element $| {\cal A}_{n+2} |^2$ is calculated perturbatively.
At leading and next-to-leading order one has the following contributions:
\bq
\l O \r^{LO} & = & 
 \int\limits_n O_n d\sigma^B, 
 \nonumber \\ 
\l O \r^{NLO} & = & 
 \int\limits_{n+1} O_{n+1} d\sigma^R + \int\limits_n O_n d\sigma^V 
 + \int\limits_n O_n d\sigma^C.
\eq
Here we used a rather condensed notation. 
$d\sigma^B$ denotes the Born contribution, while
$d\sigma^R$ denotes the real emission contribution,
whose matrix element is given by the square of the Born amplitudes with $(n+3)$ partons
$| {\cal A}^{(0)}_{n+3} |^2$.
$d\sigma^V$ gives the virtual contribution, whose matrix element is given by the interference term
of the one-loop amplitude ${\cal A}^{(1)}_{n+2}$ with $(n+2)$ partons with the corresponding
Born amplitude ${\cal A}^{(0)}_{n+2}$.
$d\sigma^C$ denotes a collinear subtraction term, which subtracts the initial-state collinear
singularities.
Within the subtraction method one constructs an approximation term $d\sigma^A$ with the same
singularity structure as $d\sigma^R$.
The NLO contribution is rewritten as
\bq
\l O \r^{NLO} & = & 
 \int\limits_{n+1} \left( O_{n+1} d\sigma^R - O_n d\sigma^A \right) 
 + \int\limits_n \left( O_n d\sigma^V + O_n d\sigma^C + O_n d\sigma^A \right),
\eq
such that the terms inside the two brackets are separately finite.
The matrix element corresponding to the approximation term $d\sigma^A$ is given as a sum over 
dipoles
\cite{Catani:1997vz,Catani:1997vzerr,Dittmaier:1999mb,Phaf:2001gc,Catani:2002hc}:
\bq
\label{eq:dipoles}
 \sum\limits_{pairs\; i,j} \;\;\; \sum\limits_{k \neq i,j} {\cal D}_{ij,k}
 + \left[
 \;\; \sum\limits_{pairs\; i,j} {\cal D}_{ij}^a
 + \sum\limits_j \sum\limits_{k \neq j} {\cal D}_{k}^{aj}
 + \sum\limits_j {\cal D}^{aj,b}
 + ( a \leftrightarrow b )
 \right].
\eq
In eq.~(\ref{eq:dipoles}) the labels $i$, $j$ and $k$ denote final-state particles, while $a$ and $b$
denote initial-state particles.
The first term describes dipoles where both the emitter and the spectator are in the final-state.
${\cal D}_{ij}^a$ denotes a dipole where the emitter is in the final-state, while
the spectator is in the initial-state.
The reverse situation is denoted by ${\cal D}_{k}^{aj}$: Here the emitter is in the initial-state
and the spectator is in the final-state.
Finally, ${\cal D}^{aj,b}$ denotes a dipole where both the emitter and the spectator are in the
initial-state.
The full complexity is only needed for hadron colliders;
for electron-positron annihilation the subtraction terms inside the
 square bracket are absent.
The dipole subtraction terms for a final-state emitter-spectator pair
have the following form:
\bq
{\cal D}_{ij,k} 
& = & 
{\cal A}_{n+2}^{(0)\;\ast}\left( p_1, ..., \tilde{p}_{(ij)},...,\tilde{p}_k,...\right)
\frac{(-{\bf T}_k \cdot {\bf T}_{ij})}{{\bf T}^2_{ij}} 
\frac{V_{ij,k}}{2 p_i \cdot p_j}
{\cal A}_{n+2}^{(0)}\left( p_1, ..., \tilde{p}_{(ij)},...,\tilde{p}_k,...\right).
 \;\;\;
\eq
The structure of the dipole subtraction terms with initial-state partons is similar. 
Here ${\bf T}_i$ denotes the colour charge operator for parton $i$ and
$V_{ij,k}$ is a matrix in the spin space of the emitter parton $(ij)$.
In general, the operators ${\bf T}_i$ lead to colour correlations, while the $V_{ij,k}$'s lead
to spin correlations.
The colour charge operators ${\bf T}_i$ for a quark, gluon and antiquark in the final state are
\bq
\label{colour_charge_operator_final}
\mbox{quark :} & & 
 {\cal A}^\ast\left(  ... q_i ... \right) \left( T_{ij}^a \right) {\cal A}\left(  ... q_j ... \right), \nonumber \\
\mbox{gluon :} & & 
 {\cal A}^\ast\left(  ... g^c ... \right) \left( i f^{cab} \right) {\cal A}\left(  ... g^b ... \right), \nonumber \\
\mbox{antiquark :} & & 
 {\cal A}^\ast\left(  ... \bar{q}_i ... \right) \left( - T_{ji}^a \right) {\cal A}\left(  ... \bar{q}_j ... \right).
\eq
The corresponding colour charge operators for a quark, gluon and antiquark in the initial state are
\bq
\label{colour_charge_operator_initial}
\mbox{quark :} & & 
 {\cal A}^\ast\left(  ... \bar{q}_i ... \right) \left( - T_{ji}^a \right) {\cal A}\left(  ... \bar{q}_j ... \right), \nonumber \\
\mbox{gluon :} & & 
 {\cal A}^\ast\left(  ... g^c ... \right) \left( i f^{cab} \right) {\cal A}\left(  ... g^b ... \right), \nonumber \\
\mbox{antiquark :} & & 
 {\cal A}^\ast\left(  ... q_i ... \right) \left( T_{ij}^a \right) {\cal A}\left(  ... q_j ... \right). 
\eq
In the amplitude an incoming quark is denoted as an outgoing antiquark and vice versa.

In this paper we neglect spin-correlations and work to leading-order in $1/N_c$.
Therefore we replace the splitting functions $V_{ij,k}$ by the spin-averaged splitting functions:
\bq
 V_{ij,k} & \rightarrow & \l V_{ij,k} \r
\eq
In the leading-colour approximation we only have to take into account emitter-spectator pairs,
which are adjacent inside a colour cluster. For those pairs we obtain for the colour
charge operators
\bq
\label{colour_average}
\frac{(-{\bf T}_k \cdot {\bf T}_{ij})}{{\bf T}^2_{ij}} 
 & = &
 \left\{
 \begin{array}{ll}
  1/2 & \mbox{emitter $(ij)$ is a gluon,} \\
  1 & \mbox{emitter $(ij)$ is a quark or antiquark.} \\
 \end{array}
 \right.
\eq
We introduce the notation
\bq
 {\cal P}_{ij,k} = \frac{\l V_{ij,k} \r}{(p_i + p_j)^2 - m_{ij}^2}
                   \cdot \theta\left(\l V_{ij,k} \r\right),
 & &
 {\cal P}_{ij,a} = \frac{\l V_{ij}^a \r}{(p_i + p_j)^2 - m_{ij}^2} \cdot \frac{1}{x} 
                   \cdot \theta\left(\l V_{ij}^a \r\right),
  \nonumber \\
 {\cal P}_{aj,k} = \frac{\l V^{aj}_k \r}{\left| 2 p_a \cdot p_j \right|} \cdot \frac{1}{x} 
                   \cdot \theta\left(\l V^{aj}_k \r\right),
  & &
 {\cal P}_{aj,b} = \frac{\l V^{aj,b} \r}{\left| 2 p_a \cdot p_j \right|} \cdot \frac{1}{x}
                   \cdot \theta\left(\l V^{aj,b} \r\right).
\eq
The functions ${\cal P}$ will govern the emission of additional particles in the shower algorithm.
The spin-averaged dipole splitting functions $\l V \r$ can be found in \cite{Catani:1997vz,Catani:2002hc}.
The Heavyside theta-functions ensure that the functions ${\cal P}$ will be non-negative.
They are needed for splittings between an initial- and a final-state particle, since
the dipole splitting functions $\l V_{ij}^a \r$ and $\l V^{aj}_k \r$ may take negative values in certain regions
of phase space.
In addition, the spin-averaged dipole splitting functions for massive partons are slightly modified:
Terms related to the soft singularity are re-arranged between the two dipoles forming an antenna,
in order to ensure positivity of the individual dipole splitting functions
in the singular limit.

% ------------------------------------------------------------------

\section{The shower algorithm}
\label{sect:shower}

In this section we describe the shower algorithm.
We first discuss the colour treatment in section~\ref{sect:colour}.
The shower algorithm for massless final-state partons is discussed in section~\ref{sect:electron_positron}.
The necessary modifications for initial-state partons are discussed in section~\ref{sect:hadron}.
Finally, massive partons are discussed in section~\ref{sect:massive}.

\subsection{Colour treatment}
\label{sect:colour}

Before starting the parton showers, the partons from the hard matrix element have to be assigned to colour clusters.
For the simplest matrix elements, like $e^+ e^- \rightarrow q \bar{q}$, the choice is unique:
The quark-antiquark pair forms a colour cluster.
For the parton shower we work in the leading-colour approximation.
In the leading-colour approximation we have to take into account only emitter-spectator pairs, which are
adjacent inside a colour cluster.
We have implemented two options:
In the first one, which we call the ``strict leading-colour approximation'', we take exactly the terms which are
leading in an expansion in $1/N_c$ and only those.
As a consequence, all splittings $g \rightarrow q \bar{q}$ are ignored, as they are colour-suppressed compared to
$g \rightarrow g g$.
In this approximation $C_F$ is replaced by
\bq 
 C_F & \rightarrow & \frac{3}{2}.
\eq
For the second option, which we call the ``modified leading-colour approximation'', we include the splitting
$g \rightarrow q \bar{q}$ and keep $C_F$ as $(N_c^2-1)/2/N_c$.
In this case, if a gluon in a closed string splits into a quark-antiquark pair, the closed string becomes an open string.
If a gluon in an open string splits into a quark-antiquark pair, the open string splits into two open strings.

\subsection{The shower algorithm for massless final-state partons}
\label{sect:electron_positron}

We first describe the shower algorithm for electron-positron annihilation.
The extension to initial-state partons is treated in section~\ref{sect:hadron}.
For the shower algorithm we use as an evolution variable
\bq
\label{evolution_variable}
 t & = & \ln \frac{-k_\perp^2}{Q^2},
\eq
where $Q^2$ is a fixed reference scale and $k_\perp$ is the transverse momentum of a splitting.
During the shower evolution we move towards
smaller (more negative) values of $t$.
We start from a given $n$-parton configuration.
In the dipole formalism, emission of additional partons is described by an 
emitter-spectator pair.
In the leading colour approximation emitter and spectator are always adjacent 
in the cyclic order.
The probability to evolve from $t_1$ to $t_2$ (with $t_1 > t_2$) without any resolvable branching is given by the Sudakov factor.
For the algorithm considered here, the Sudakov factor is given 
as a product of factors corresponding 
to the no-emission probabilities for individual dipoles' emissions:
\bq
\label{overall_sudakov}
 \Delta(t_1,t_2) & = &
  \prod\limits_{\tilde{i},\tilde{k}} \Delta_{\tilde{i},\tilde{k}}(t_1,t_2).
\eq
If parton $\tilde{i}$ can emit different partons, 
$\Delta_{\tilde{i},\tilde{k}}(t_1,t_2)$ factorises in turn into different contributions:
\bq
 \Delta_{\tilde{i},\tilde{k}}(t_1,t_2) & = & \prod\limits_{j} \Delta_{ij,k}(t_1,t_2),
\eq
An example is the possibility of a gluon to split either into two gluons or into a $\bar{q} q$-pair.
We denote the emitter before the splitting by $\tilde{i}$, while the emitter after a splitting is denoted
by $i$.
This notation takes into account that the emitter might change its ``flavour'' due to a splitting, like
in the case of a $g \rightarrow \bar{q} q$ splitting.
$\Delta_{ij,k}(t_1,t_2)$ is the probability that the dipole formed by the emitter
$\tilde{i}$ and spectator $\tilde{k}$ does not emit a parton $j$.
It is given by
\bq
\label{individual_sudakov}
 \Delta_{ij,k}(t_1,t_2) & = &
 \exp\left( - \int\limits_{t_2}^{t_1} dt {\cal C}_{\tilde{i},\tilde{k}} 
              \int d\phi_{unres} \delta\left(t-T_{\tilde{i},\tilde{k}} \right) {\cal P}_{ij,k} \right),
\eq
where ${\cal C}_{\tilde{i},\tilde{k}}$ is  a colour factor.
In the leading colour approximation this factor is non-zero only if $\tilde{i}$ and $\tilde{k}$
are adjacent in a colour cluster. Then ${\cal C}_{\tilde{i},\tilde{k}}$ is obtained from
eq.~(\ref{colour_average}) and given by
\bq
 C_{\tilde{i},\tilde{k}} & = & \left\{ 
 \begin{array}{lll}
  \frac{1}{2} & \mbox{for} & \tilde{i}=g, \\
  1 & \mbox{for} & \tilde{i} = q, \bar{q}. \\
 \end{array}
\right.
\eq
The dipole phase space is given by
\bq
 \int d\phi_{unres} & = & 
  \frac{(p_{\tilde{i}}+p_{\tilde{k}})^2}{16\pi^2} \int\limits_0^1 d\kappa \int\limits_{z_-(\kappa)}^{z_+(\kappa)} dz
     \frac{1}{4z(1-z)} \left( 1 - \frac{\kappa}{4 z (1-z)} \right),
\eq
with
\bq 
 z_{\pm}(\kappa) & = & \frac{1}{2} \left( 1 \pm \sqrt{1-\kappa} \right).
\eq
The variable $\kappa$ is proportional to the transverse momentum of the splitting
\bq
 \kappa & = & 4 \frac{(-k_\perp^2)}{(p_{\tilde{i}}+p_{\tilde{k}})^2}.
\eq
$T_{\tilde{i},\tilde{k}}$ depends on the dipole invariant mass $(p_{\tilde{i}}+p_{\tilde{k}})^2$ and the
phase space variable $\kappa$ for the emission of an additional particle and is given by
\bq
 T_{\tilde{i},\tilde{k}} & = & 
 \ln \frac{\kappa}{4} \frac{(p_{\tilde{i}}+p_{\tilde{k}})^2}{Q^2}
\eq
With the help of the delta-function we may perform the integration over $\kappa$, while keeping the
integration over $t$ and $z$.  Then
\bq
 \kappa(t) & = &
 \frac{4Q^2 e^t}{(p_{\tilde{i}}+p_{\tilde{k}})^2}.
\eq
${\cal P}_{ij,k}$ is the dipole splitting function.
As an example we quote the splitting function for the $q\rightarrow q g$ splitting:
\bq
{\cal P}_{q \rightarrow q g} = 
   C_F
   \frac{8\pi \alpha_s(\mu^2)}{(p_{\tilde{i}}+p_{\tilde{k}})^2} \frac{1}{y}
   \left[ \frac{2}{1-z(1-y)} - (1+z) \right],
& &
 y = \frac{\kappa(t)}{4 z (1-z)}.
\eq
$\alpha_s$ is evaluated at the scale $\mu^2=-k_\perp^2=\frac{\kappa}{4} (p_{\tilde{i}}+p_{\tilde{k}})^2$.
The probability that a branching occurs at $t_2$ is given by
\bq
  \sum\limits_{\tilde{i},\tilde{k}} \;\;\; \sum\limits_{j} 
   {\cal C}_{\tilde{i},\tilde{k}} \int d\phi_{unres} \delta\left(t_2-T_{\tilde{i},\tilde{k}} \right) {\cal P}_{ij,k}
  \Delta(t_1,t_2).
\eq
We can now state the shower algorithm. Starting from an initial evolution scale $t_1$ we proceed as follows:
\begin{enumerate}

\item Select the next dipole to branch and the scale $t_2$ at which this occurs.
This is done as follows:
For each dipole we generate the scale $t_{2,ij,k}$ of the next splitting for this dipole
from a uniformly distributed
number $r_{1,ij,k}$ in $[0,1]$ by solving (numerically) the equation
\bq
 \Delta_{ij,k}(t_1,t_{2,ij,k}) & = & r_{1,ij,k}.
\eq
We then set
\bq 
t_2 & = & \mbox{max} \left( t_{2,ij,k} \right).
\eq
The dipole which has the maximal value of $t_{2,ij,k}$ is the one which radiates off an additional
particle.

\item If $t_2$ is smaller than a cut-off scale $t_{min}$, the shower algorithm terminates.

\item 
Next we have to generate the value of $z$. Again, using a uniformly distributed random number $r_2$ in $[0,1]$ we solve
\bq
\label{z_selection}
 \int\limits_{z_-(t_2)}^{z} dz' J(t_2,z') {\cal P}_{ij,k}
 & = & r_2 \int\limits_{z_-(t_2)}^{z_+(t_2)} dz' J(t_2,z') {\cal P}_{ij,k},
\eq
where the Jacobian factor $J(t_2,z)$ is given by
\bq
 J(t_2,z) & = & 
  \frac{\kappa(t_2)}{4z(1-z)} \left( 1 - \frac{\kappa(t_2)}{4 z (1-z)} \right).
\eq

\item Select the azimuthal angle $\phi$.
Finally we generate the azimuthal angle from a uniformly distributed number $r_3$ in $[0,1]$ as follows:
\bq
 \phi & = & 2 \pi r_3.
\eq

\item With the three kinematical variables $t_2$, $z$ and $\phi$ and the information, that
parton $\tilde{i}$ emits a parton $j$, with parton $\tilde{k}$ being the spectator, we insert the new parton
$j$. The momenta $p_{\tilde{i}}$ and $p_{\tilde{k}}$ of the emitter and the spectator are replaced by new momenta
$p_i$ and $p_k$.
The details how the new momenta $p_i$, $p_j$ and $p_k$ are constructed are given in the appendix~\ref{sect:insertion}.

\item Set $t_1=t_2$ and go to step 1.

\end{enumerate}

Remark: Step 1 of the algorithm is equivalent to first
generating the point $t_2$ from a uniformly distributed
number $r_1$ in $[0,1]$ by solving (numerically) the equation for the full Sudakov factor
\bq
 \Delta(t_1,t_2) & = & r_1,
\eq
and then selecting an individual dipole 
with emitter $\tilde{i}$, emitted particle $j$ and spectator
$k$ with probability \cite{Seymour:1995df} 
\bq
\label{dipole_selection}
 P_{ij,k} & = &
   \frac{
    {\cal C}_{\tilde{i},\tilde{k}} \int d\phi_{unres} \delta\left(t_2-T_{\tilde{i},\tilde{k}} \right) {\cal P}_{ij,k}
 }{
  \sum\limits_{\tilde{l},\tilde{n}} \;\;\; \sum\limits_{m} 
   {\cal C}_{\tilde{l},\tilde{n}} \int d\phi_{unres} \delta\left(t_2-T_{\tilde{l},\tilde{n}} \right) {\cal P}_{lm,n}
}.
\eq

\subsection{The shower algorithm with initial-state partons}
\label{sect:hadron}

In this subsection we discuss the necessary modifications 
for the inclusion of initial-state partons.
In the presence of initial-state partons there is no separation into final-state showers and initial-state
showers.
Initial-state radiation is treated on the same footing as final-state radiation.
The algorithm generates initial-state radiation through backward evolution, starting from 
a hard scale and moving towards softer scales.
Therefore the shower evolves in all cases from a hard scale towards lower scales.

\subsubsection*{Final-state emitter and initial-state spectator}

For an initial-state spectator we modify the Sudakov factor in eq.~(\ref{individual_sudakov})
to
\bq
 \Delta_{ij,a}(t_1,t_2) & = &
 \exp\left( - \int\limits_{t_2}^{t_1} dt {\cal C}_{\tilde{i},\tilde{a}} 
              \int d\phi_{unres} \delta\left(t-T_{\tilde{i},\tilde{a}} \right) 
                                 \frac{x_a f(x_a,t)}{x_{\tilde{a}} f(x_{\tilde{a}},t)}
                                 {\cal P}_{ij,a} 
     \right),
\eq
where $x_{\tilde{a}}$ is the momentum fraction of the initial hadron carried by $\tilde{a}$, while
$x_a$ is the momentum fraction carried by $a$.
The initial parton of the $n$-particle state is denoted by $\tilde{a}$, while the initial parton
of the $(n+1)$-particle state is denoted by $a$.
We set
\bq
 x & = & \frac{x_{\tilde{a}}}{x_a}.
\eq
The unresolved phase space is given by
\bq
\label{initial_final_unres_ps}
 \int d\phi_{unres} & = & 
 \frac{\left| 2p_{\tilde{i}}p_{\tilde{a}} \right|}{16 \pi^2} 
 \int\limits_{x_{\tilde{a}}}^1 \frac{dx}{x} 
 \int\limits_0^1 dz.
\eq
The transverse momentum between $i$ and $j$ is expressed as
\bq
 - k_\perp^2 & = & \frac{(1-x)}{x} z (1-z) \left(-2p_{\tilde{i}}p_{\tilde{a}}\right)
\eq
and 
$T_{\tilde{i},\tilde{a}}$ is therefore given by
\bq
 T_{\tilde{i},\tilde{a}} & = & 
  \ln \frac{-k_\perp^2}{Q^2}
 =
  \ln \frac{\left(-2p_{\tilde{i}}p_{\tilde{a}}\right)(1-x) z(1-z)}{x Q^2}.
\eq
A subtlety occurs for the emission between a final-state spectator and an initial-state emitter.
We discuss this for the splitting $q\rightarrow qg$.
The spin-averaged splitting function for the $q\rightarrow qg$ splitting is given by
\bq
 \l V_{qg}^k \r & = & 8 \pi \alpha_s \; C_F
 \left[ \frac{2}{1-z+(1-x)} - (1+z) \right].
\eq
In contrast to the final-final case this function is not a positive function on the complete
phase-space. It can take negative values in certain (non-singular) regions of phase-space.
This is no problem for its use as a subtraction terms in NLO calculations, but prohibits
a straightforward interpretation as a splitting probability for a shower algorithm.
However, since negative values occur only in non-singular regions, we can ensure positiveness
by modifying the splitting functions through non-singular terms.
The simplest choice is to set
\bq
 {\cal P}_{ij,a} & = & \frac{\l V_{ij}^a \r}{(p_i + p_j)^2} \cdot \frac{1}{x} 
                       \cdot \theta\left(\l V_{ij}^a \r\right).
\eq
For a final-state emitter we eliminate the $x$-integration with the help of the delta-function:
\bq
 \int\limits_{x_{\tilde{a}}}^1 \frac{dx}{x} \delta\left(t-T_{\tilde{i},\tilde{a}} \right)
  = 
  \frac{1}{1+\frac{4 z (1-z)}{\kappa(t)} },
 & &
 x = \frac{1}{1+\frac{\kappa(t)}{4z(1-z)}},
 \;\;\;
 \kappa(t) = \frac{4 Q^2 e^t}{\left(-2p_{\tilde{i}}p_{\tilde{a}}\right)}.
\eq
For the boundaries we obtain
\bq
 \kappa(t) < \frac{1-x_{\tilde{a}}}{{x_{\tilde{a}}}},
 \;\;\;
 z_-(t) < z < z_+(t),
 \;\;\;
 z_{\pm}(t) = \frac{1}{2} \left( 1 \pm \sqrt{1-\kappa(t) \frac{{x_{\tilde{a}}}}{1-{x_{\tilde{a}}}}} \right).
\eq
The modifications to the shower algorithm are as follows:
The dipoles for the emission from a final-state emitter with an initial-state spectator are included
in the Sudakov factor in eq.~(\ref{overall_sudakov}).
With this modification steps 1 and 2 are as above.
Let us define
\bq
 f_{lm,n}
 & = & 
 \left\{ \begin{array}{ll}
 1, & \mbox{if $l$ and $n$ are final-state particles}, \\
 \frac{x_a f(x_a,t)}{x_{\tilde{a}} f(x_{\tilde{a}},t)}, & \mbox{if $l=a$ is an initial-state particle}, \\
 \frac{x_b f(x_b,t)}{x_{\tilde{b}} f(x_{\tilde{b}},t)}, & \mbox{if $n=b$ is an initial-state particle and $l$ is a final-state particle}. \\
 \end{array} \right.
\eq
In step 2 we replace formula~(\ref{z_selection}) by
\bq
 \int\limits_{z_-(t_2)}^{z} dz' J(t_2,z') f_{ij,a} {\cal P}_{ij,a}
 & = & r_2 \int\limits_{z_-(t_2)}^{z_+(t_2)} dz' J(t_2,z') f_{ij,a} {\cal P}_{ij,a},
\eq
with the Jacobian
\bq
 J(t,z) & = &  \frac{1}{1+\frac{4 z (1-z)}{\kappa(t)} }.
\eq
Steps 4 to 6 proceed as in the case described above.

\subsubsection*{Initial-state emitter and final-state spectator}

For an initial-state emitter $\tilde{a}$ with a final-state spectator $\tilde{i}$ 
the Sudakov factor is given by
\bq
 \Delta_{aj,i}(t_1,t_2) & = &
 \exp\left( - \int\limits_{t_2}^{t_1} dt {\cal C}_{\tilde{a},\tilde{i}} 
              \int d\phi_{unres} \delta\left(t-T_{\tilde{a},\tilde{i}} \right) 
                                 \frac{x_a f(x_a,t)}{x_{\tilde{a}} f(x_{\tilde{a}},t)}
                                 {\cal P}_{aj,i} 
     \right).
\eq
The unresolved phase space is again given by eq.~(\ref{initial_final_unres_ps}).
The transverse momentum between $a$ and $j$ is given by
\bq
 - k_\perp^2 & = & \frac{(1-x)}{x} (1-z) \left(-2p_{\tilde{i}}p_{\tilde{a}}\right)
\eq
and
$T_{\tilde{a},\tilde{i}}$ is given by
\bq
 T_{\tilde{a},\tilde{i}} & = &
  \ln \frac{\left(-2p_{\tilde{i}}p_{\tilde{a}}\right)(1-x) (1-z)}{x Q^2}.
\eq
For a initial-state emitter we eliminate the $z$-integration with the help of the delta-function:
\bq
 \int\limits_0^1 dz \delta\left(t-T_{\tilde{a},\tilde{i}} \right)
  =
  \frac{\kappa(t)}{4} \frac{x}{(1-x)},
 & &
 z = 1 - \frac{\kappa(t)}{4} \frac{x}{(1-x)},
 \;\;\;
 \kappa(t) = \frac{4 Q^2 e^t}{\left(-2p_{\tilde{i}}p_{\tilde{a}}\right)}.
\eq
For the boundaries we obtain
\bq
 \kappa(t) < 4 \frac{1-x_{\tilde{a}}}{x_{\tilde{a}}},
 \;\;\;
 x < x_+(t),
 \;\;\;
 x_+(t) = \frac{1}{1+\frac{\kappa(t)}{4}}.
\eq
There are no new modifications to the shower algorithms compared to the case for a final-state emitter 
and an initial-state spectator, except that in step 3 we now generate the value of $x$ according to
\bq
 \int\limits_{x_{\tilde{a}}}^{x} dx' J(t_2,x') f_{aj,i} {\cal P}_{aj,i}
 & = & r_2 \int\limits_{x_{\tilde{a}}}^{x_+(t_2)} dx' J(t_2,x') f_{aj,i} {\cal P}_{aj,i},
\eq
with the Jacobian
\bq
 J(t,x) & = & \frac{\kappa(t)}{4(1-x)}.
\eq

\subsubsection*{Initial-state emitter and initial-state spectator}

For an initial-state emitter $\tilde{a}$ with an initial-state spectator $\tilde{b}$ 
the Sudakov factor is given by
\bq
 \Delta_{aj,b}(t_1,t_2) & = &
 \exp\left( - \int\limits_{t_2}^{t_1} dt {\cal C}_{\tilde{a},\tilde{b}} 
              \int d\phi_{unres} \delta\left(t-T_{\tilde{a},\tilde{b}} \right) 
                                 \frac{x_a f(x_a,t)}{x_{\tilde{a}} f(x_{\tilde{a}},t)}
                                 {\cal P}_{aj,b} 
     \right).
\eq
In this case we do not rescale the momentum of the spectator, but transform all final-state momenta.
Therefore no factor
\bq
 \frac{x_b f(x_b,t)}{x_{\tilde{b}} f(x_{\tilde{b}},t)}
\eq
appears in the Sudakov factor.
The unresolved phase space is given by
\bq
\label{initial_initial_unres_ps}
 \int d\phi_{unres} & = & 
 \frac{\left| 2p_{\tilde{a}}p_{\tilde{b}} \right|}{16 \pi^2} 
 \int\limits_{x_{\tilde{a}}}^1 \frac{dx}{x} 
 \int\limits_0^{1-x} dv.
\eq
The transverse momentum between $a$ and $j$ is given by
\bq
 - k_\perp^2 & = & \frac{(1-x)}{x} v \left(2p_{\tilde{a}}p_{\tilde{b}}\right)
\eq
and
$T_{\tilde{a},\tilde{b}}$ is given by
\bq
 T_{\tilde{a},\tilde{b}} & = &
  \ln \frac{\left(2p_{\tilde{a}}p_{\tilde{b}}\right)(1-x) v}{x Q^2}.
\eq
We integrate over $v$ with the help of the delta-function:
\bq
 \int\limits_0^{1-x} dv \delta\left(t-T_{\tilde{a},\tilde{b}} \right)
  =
  \frac{\kappa(t)}{4} \frac{x}{(1-x)},
 & &
  v = \frac{\kappa(t)}{4} \frac{x}{(1-x)},
 \;\;\;
 \kappa(t) = \frac{4 Q^2 e^t}{\left(2p_{\tilde{a}}p_{\tilde{b}}\right)}.
\eq
For the boundaries we obtain
\bq
 \kappa(t) < 4 \frac{(1-x_{\tilde{a}})^2}{x_{\tilde{a}}},
 \;\;\;
 x < x_+(t),
 \;\;\;
 x_+(t) = \frac{1}{2} \left( 2 + \frac{\kappa(t)}{4} - \sqrt{\kappa(t) + \frac{\kappa(t)^2}{16}} \right).
\eq
In step 3 of the shower algorithm we again select $x$ according to
\bq
 \int\limits_{x_{\tilde{a}}}^{x} dx' J(t_2,x') f_{aj,b} {\cal P}_{aj,b}
 & = & r_2 \int\limits_{x_{\tilde{a}}}^{x_+(t_2)} dx' J(t_2,x') f_{aj,b} {\cal P}_{aj,b},
\eq
with the Jacobian
\bq
 J(t,x) & = & \frac{\kappa(t)}{4(1-x)}.
\eq

\subsection{The shower algorithm for massive partons}
\label{sect:massive}

In this subsection we discuss the modifications of the shower
algorithms due to the presence of massive partons.
We first address the issue of a splitting of a gluon into a heavy quark pair.
This mainly concerns the splitting of a gluon into $b$-quarks.
We will always require that initial-state particles are massless.
Therefore for processes with initial-state hadrons we do not consider
$g \rightarrow Q \bar{Q}$ splittings.
Calculations for initial-state hadrons should be done in the approximation of a massless
$b$-quark.
In the case of electron-positron annihilation the parton shower affects only the final state.
Here we can consistently allow splittings of a gluon into a pair of massive quarks.
As evolution variable we use in the massive case
\bq
 t & = & \ln \frac{-k_\perp^2+(1-z)^2 m_i^2 + z^2 m_j^2}{Q^2}.
\eq
This choice reduces to eq.~(\ref{evolution_variable}) in the massless limit 
and is suggested by dispersion relations for
the running coupling \cite{Dokshitzer:1995qm,Gieseke:2003rz}.

\subsubsection*{Final-state emitter and final-state spectator}

The unresolved phase space is given by
\bq
\label{massive_final_final_unres_ps}
 \int d\phi_{unres} & = & 
  \frac{(p_{\tilde{i}}+p_{\tilde{k}})^2}{16\pi^2} 
  \left( 1 - \mu_i^2 - \mu_j^2 - \mu_k^2 \right)^2
  \left[ \lambda(1,\mu_{ij}^2,\mu_k^2) \right]^{-\frac{1}{2}}
  \int\limits_{y_-}^{y_+} dy \; (1-y) \int\limits_{z_-(y)}^{z_+(y)} dz,
\eq
where the reduced masses $\mu_l$ and the boundaries on the integrations are defined in appendix \ref{sect:insertion}
in eqs. (\ref{def_massive_1})-(\ref{def_massive_4}).
$T_{\tilde{i},\tilde{k}}$ is given by
\bq
 T_{\tilde{i},\tilde{k}} & = & 
 \ln \frac{\left( (p_{\tilde{i}}+p_{\tilde{k}})^2 -m_i^2-m_j^2-m_k^2\right)y z (1-z)}{Q^2}.
\eq
Again, we have to ensure that the splitting functions are positive.
The original spin-averaged dipole splitting functions can take negative values in certain regions of phase-space.
In the massive case the negative region can extend into the singular region.
The problem is related to the soft behaviour of the dipole splitting functions.
Since a squared Born matrix element is positive in the soft gluon limit,
the negative contribution from a particular dipole is compensated by
the contribution from the dipole, where emitter and spectator are
exchanged. The sum of the two contributions is positive in the singular region.
Therefore we can cut out the negative region from the first dipole and add it to the
second dipole. The second dipole will stay positive.

As in the massless case we eliminate the $y$-integration:
\bq
  \int\limits_{y_-}^{y_+} dy \; (1-y) \int\limits_{z_-(y)}^{z_+(y)} dz 
   \delta\left(t-T_{\tilde{i},\tilde{k}}\right)
& = & 
 \int\limits_{z_{min}}^{z_{max}} dz \; y(1-y),
 \\
& &
 y = \frac{\kappa(t)}{4z(1-z)},
 \;\;\;
 \kappa(t) = \frac{4 Q^2 e^t}{(p_{\tilde{i}}+p_{\tilde{k}})^2 -m_i^2-m_j^2-m_k^2}.
 \nonumber
\eq
The physical region is defined by
\bq
 \left( 1- \frac{\kappa}{4z(1-z)} \right)^2
  \left[ \frac{\kappa}{4} - (1-z)^2 \bar{m}_i^2 - z^2 \bar{m}_j^2 \right]
 - \left( \frac{\kappa}{4z(1-z)} \right)^2 \bar{m}_k^2
 + 4 \bar{m}_i^2 \bar{m}_j^2 \bar{m}_k^2 
 & \ge & 0,
\eq
with
\bq
 \bar{m}_l^2 & = & \frac{m_l^2}{(p_{\tilde{i}}+p_{\tilde{k}})^2 -m_i^2-m_j^2-m_k^2}
\;\;\;\mbox{for}\;\;\; l \in \{i,j,k\}.
\eq
This equation is solved numerically for $z_{min}$ and $z_{max}$.
Then $z$ is generated according to
\bq
 \int\limits_{z_{min}(t_2)}^{z} dz' J(t_2,z') {\cal P}_{ij,k}
 & = & r_2 \int\limits_{z_{min}(t_2)}^{z_{max}(t_2)} dz' J(t_2,z') {\cal P}_{ij,k},
\eq
with the Jacobian
\bq
 J(t,z) & = & 
  \left( 1 - \mu_i^2 - \mu_j^2 - \mu_k^2 \right)^2
  \left[ \lambda(1,\mu_{ij}^2,\mu_k^2) \right]^{-\frac{1}{2}}
  \frac{\kappa(t)}{4z(1-z)} \left( 1 - \frac{\kappa(t)}{4 z (1-z)} \right).
\eq

\subsubsection*{Final-state emitter and initial-state spectator}

The unresolved phase space is given by
\bq
\label{massive_initial_final_unres_ps}
 \int d\phi_{unres} & = & 
 \frac{\left| 2p_{\tilde{i}}p_{\tilde{a}} \right|}{16 \pi^2} 
 \int\limits_{x_{\tilde{a}}}^1 \frac{dx}{x} 
 \int\limits_{z_-(x)}^{1} dz
 =
 \frac{\left| 2p_{\tilde{i}}p_{\tilde{a}} \right|}{16 \pi^2} 
 \int\limits_{z_-(x_{\tilde{a}})}^{1} dz
 \int\limits_{x_{\tilde{a}}}^{x_+(z)} \frac{dx}{x} 
\eq
The integration boundary is given by
\bq
 z_-(x) = \frac{x\tilde{\mu}^2}{1-x(1-\tilde{\mu}^2)},
 \;\;\;
 x_+(z) = \frac{z}{\tilde{\mu}^2+z(1-\tilde{\mu}^2)},
 \;\;\;
 \tilde{\mu}^2 = \frac{m_i^2}{\left| 2 p_{\tilde{i}} p_{\tilde{a}} \right|}.
\eq
$T_{\tilde{i},\tilde{a}}$ is given by
\bq
 T_{\tilde{i},\tilde{a}} & = & 
  \ln \frac{-k_\perp^2+(1-z)^2m_i^2}{Q^2}
 =
  \ln \frac{\left(-2p_{\tilde{i}}p_{\tilde{a}}\right)(1-x) z(1-z)}{x Q^2}.
\eq
For a final-state emitter we eliminate the $x$-integration with the help of the delta-function:
\bq
 \int\limits_{x_{\tilde{a}}}^1 \frac{dx}{x} \delta\left(t-T_{\tilde{i},\tilde{a}} \right)
  = 
  \frac{1}{1+\frac{4 z (1-z)}{\kappa(t)} },
 & &
 x = \frac{1}{1+\frac{\kappa(t)}{4z(1-z)}},
 \;\;\;
 \kappa(t) = \frac{4 Q^2 e^t}{\left(-2p_{\tilde{i}}p_{\tilde{a}}\right)}.
\eq
For the boundaries we obtain
\bq
z_+(t) & = & \frac{1}{2} \left( 1 + \sqrt{ 1 - \kappa(t) \frac{x_{\tilde{a}}}{1-x_{\tilde{a}}}} \right),
 \nonumber \\
z_-(t) & = &
 \mbox{max}\left( 
  \frac{x_{\tilde{a}}\tilde{\mu}^2}{1-x_{\tilde{a}}\left(1-\tilde{\mu}^2\right)},
  \frac{1}{2} \left( 1 - \sqrt{ 1 - \kappa(t) \frac{x_{\tilde{a}}}{1-x_{\tilde{a}}}} \right),
  1 - \sqrt{\frac{\kappa(t)}{4\tilde{\mu}^2}}
 \right).
\eq
The boundary on $\kappa(t)$ is given for $\tilde{\mu}^2< (1-x_{\tilde{a}})/x_{\tilde{a}}$ by
\bq
 \kappa(t) & < & \frac{1-x_{\tilde{a}}}{x_{\tilde{a}}}.
\eq
For $(1-x_{\tilde{a}})/x_{\tilde{a}} < \tilde{\mu}^2$ we have
\bq
\kappa(t) & < & 
 \frac{1-x_{\tilde{a}}}{x_{\tilde{a}}}
 \left[ 1 - \left( \frac{1-\frac{1-x_{\tilde{a}}}{x_{\tilde{a}} \tilde{\mu}^2} }
                        {1+\frac{1-x_{\tilde{a}}}{x_{\tilde{a}} \tilde{\mu}^2} } \right)^2
 \right]
 =
 \frac{4\tilde{\mu}^2}{\left(1+\frac{x_{\tilde{a}}\tilde{\mu}^2}{1-x_{\tilde{a}}}\right)^2}.
\eq
$z$ is generated according to
\bq
 \int\limits_{z_-(t_2)}^{z} dz' J(t_2,z') f_{ij,a} {\cal P}_{ij,a}
 & = & r_2 \int\limits_{z_-(t_2)}^{z_+(t_2)} dz' J(t_2,z') f_{ij,a} {\cal P}_{ij,a},
\eq
with the Jacobian 
\bq
 J(t,z) & = & \frac{1}{1+\frac{4 z (1-z)}{\kappa(t)} }.
\eq

\subsubsection*{Initial-state emitter and final-state spectator}

$T_{\tilde{a},\tilde{i}}$ is given by
\bq
 T_{\tilde{a},\tilde{i}} & = &
  \ln \frac{\left(-2p_{\tilde{i}}p_{\tilde{a}}\right)(1-x) (1-z)}{x Q^2}.
\eq
For an initial-state emitter we eliminate the $z$-integration with the help of the delta-function:
\bq
 \int\limits_{z_-(x)}^1 dz \delta\left(t-T_{\tilde{a},\tilde{i}} \right)
  =
  \frac{\kappa(t)}{4} \frac{x}{(1-x)},
 & &
 z = 1 - \frac{\kappa(t)}{4} \frac{x}{(1-x)},
 \;\;\;
 \kappa(t) = \frac{4 Q^2 e^t}{\left(-2p_{\tilde{i}}p_{\tilde{a}}\right)}.
\eq
For the boundaries we obtain
\bq
 \kappa(t) < 
 \frac{4\left(1-x_{\tilde{a}}\right)^2}{x_{\tilde{a}}\left[1-x_{\tilde{a}}\left(1-\tilde{\mu}^2\right)\right]},
\;\;\
 x < x_+(t),
\;\;\;
 x_+(t) =  
 \frac{ 2 + \frac{\kappa(t)}{4} 
        - \sqrt{ \frac{\kappa(t)^2}{16} + \tilde{\mu}^2 \kappa(t) } }
      {2 \left( 1+\frac{\kappa(t)}{4}\left(1-\tilde{\mu}^2\right) \right)}, 
\eq
The value of $x$ is generated according to
\bq
 \int\limits_{x_{\tilde{a}}}^{x} dx' J(t_2,x') f_{aj,i} {\cal P}_{aj,i}
 & = & r_2 \int\limits_{x_{\tilde{a}}}^{x_+(t_2)} dx' J(t_2,x') f_{aj,i} {\cal P}_{aj,i},
\eq
with the Jacobian
\bq
 J(t,x) & = & \frac{\kappa(t)}{4(1-x)}.
\eq

% ------------------------------------------------------------------

\section{Numerical results}
\label{sect:results}

In this section we show numerical results 
obtained from the parton shower.
We first discuss in section~\ref{sect:results_electron_positron}
observables related to electron-positron annihilation
and then in section~\ref{sect:results_hadron} the shower in hadron
collisions.
The shower algorithm depends on two parameters, the strong coupling $\alpha_s$ and the
scale $Q_{min}$.
For the strong coupling we use the leading-order formula
\bq
 \alpha_s(\mu) & = & \frac{4\pi}{\beta_0 \ln \frac{\mu^2}{\Lambda^2}},
 \;\;\;
 \beta_0 = 11 - \frac{2}{3} N_f.
\eq
The cut-off scale $Q_{min}$ gives the scale at which the shower terminates.
As our shower is correct in the leading-colour approximation, we also study the effects
of different treatments of sub-leading colour contributions.
As described in section~\ref{sect:colour} we have implemented two options:
The strict leading-colour approximation and the modified leading-colour approximation.
Numerical differences from these two options will give an estimate of uncertainties 
due to subleading-colour effects.

\subsection{Electron-positron annihilation}
\label{sect:results_electron_positron}

\begin{figure}[ht]
\begin{center}
\includegraphics[bb= 120 460 500 730,width=0.5\textwidth]{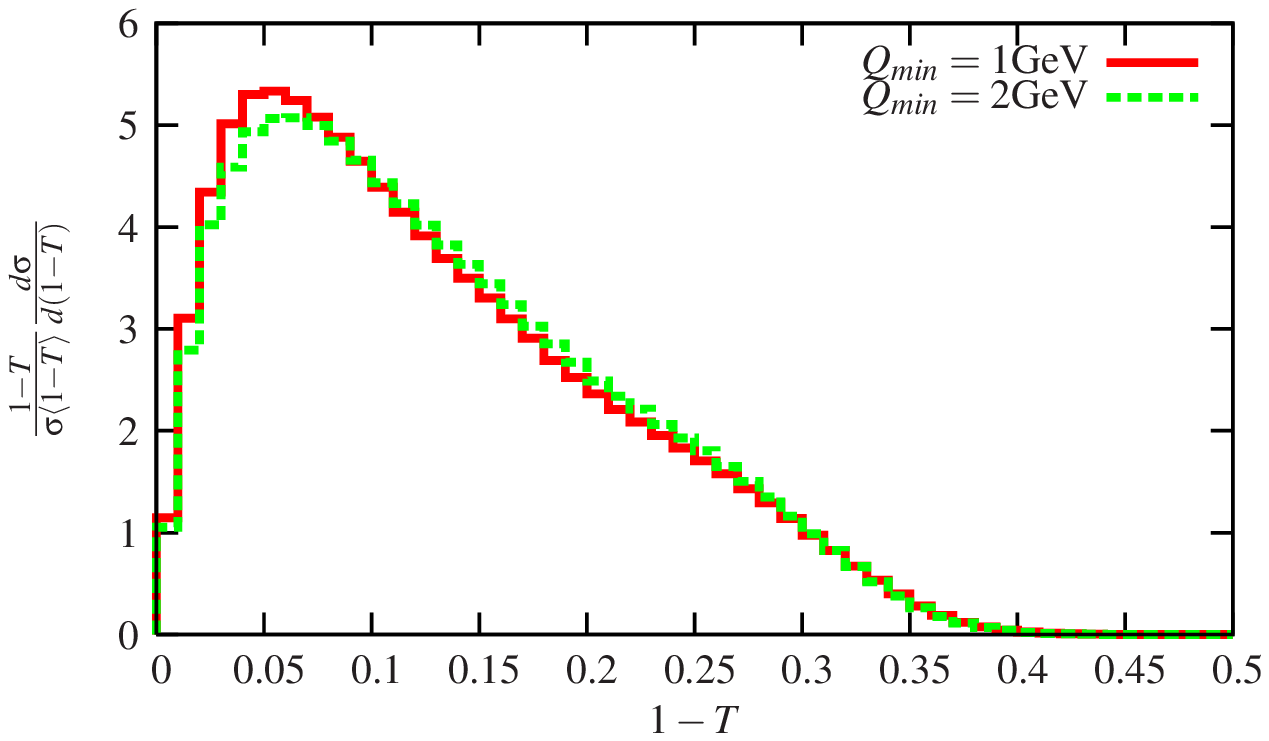}
\includegraphics[bb= 120 460 500 730,width=0.5\textwidth]{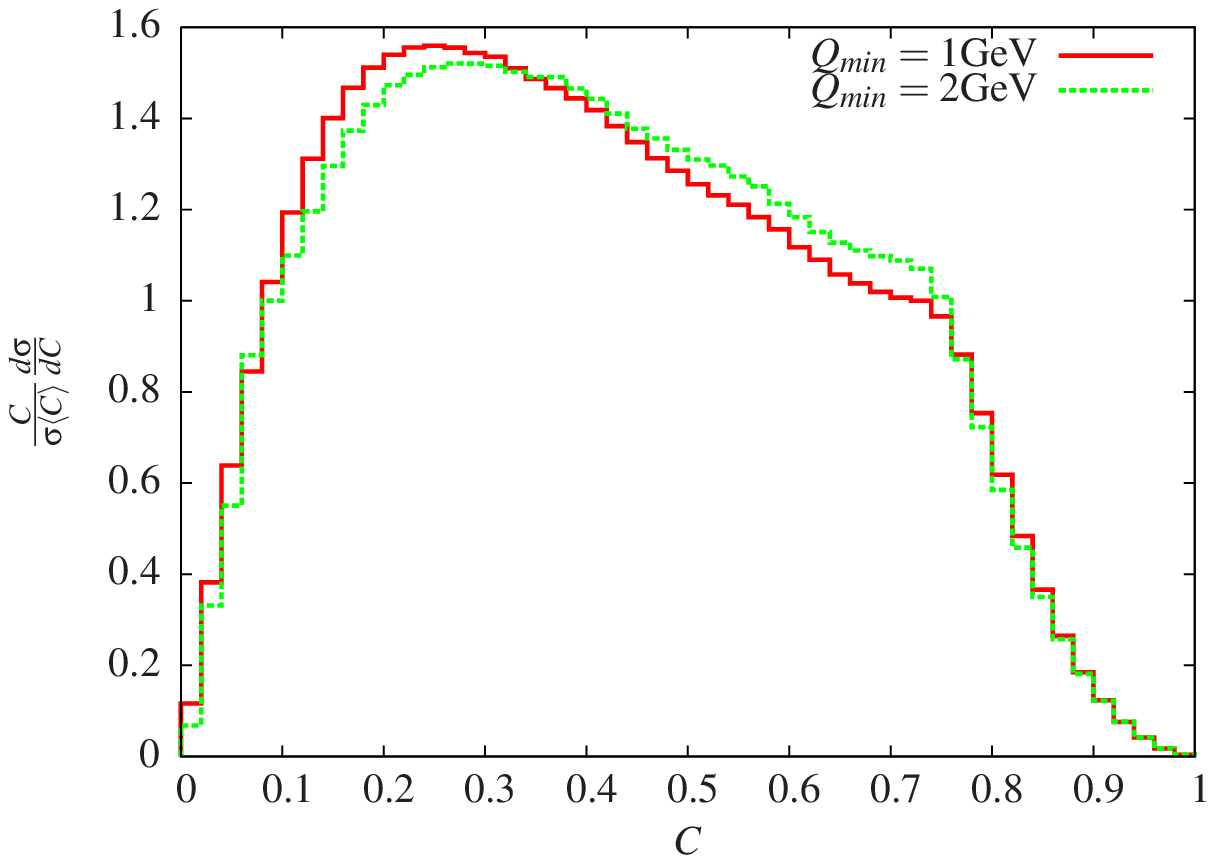}
\includegraphics[bb= 120 460 500 730,width=0.5\textwidth]{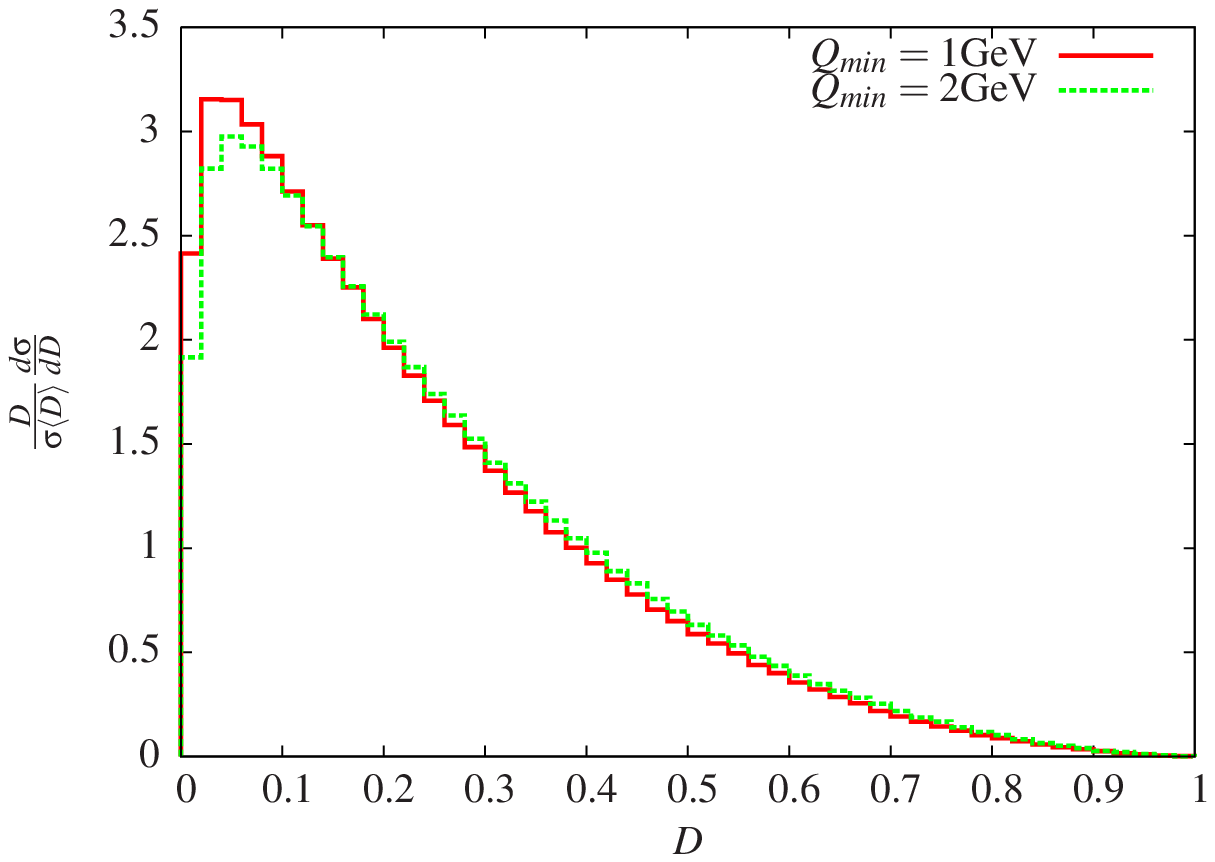}
\end{center}
\caption{The first moments of the thrust distribution, the C-parameter distribution and the D-parameter distribution.
The results are from the parton shower for two different values of the cut-off scale $Q_{min}$.}
\label{fig_thrust}
\end{figure}

\begin{figure}[ht]
\begin{center}
\begin{tabular}{ll}
\includegraphics[bb= 120 460 500 730,width=0.5\textwidth]{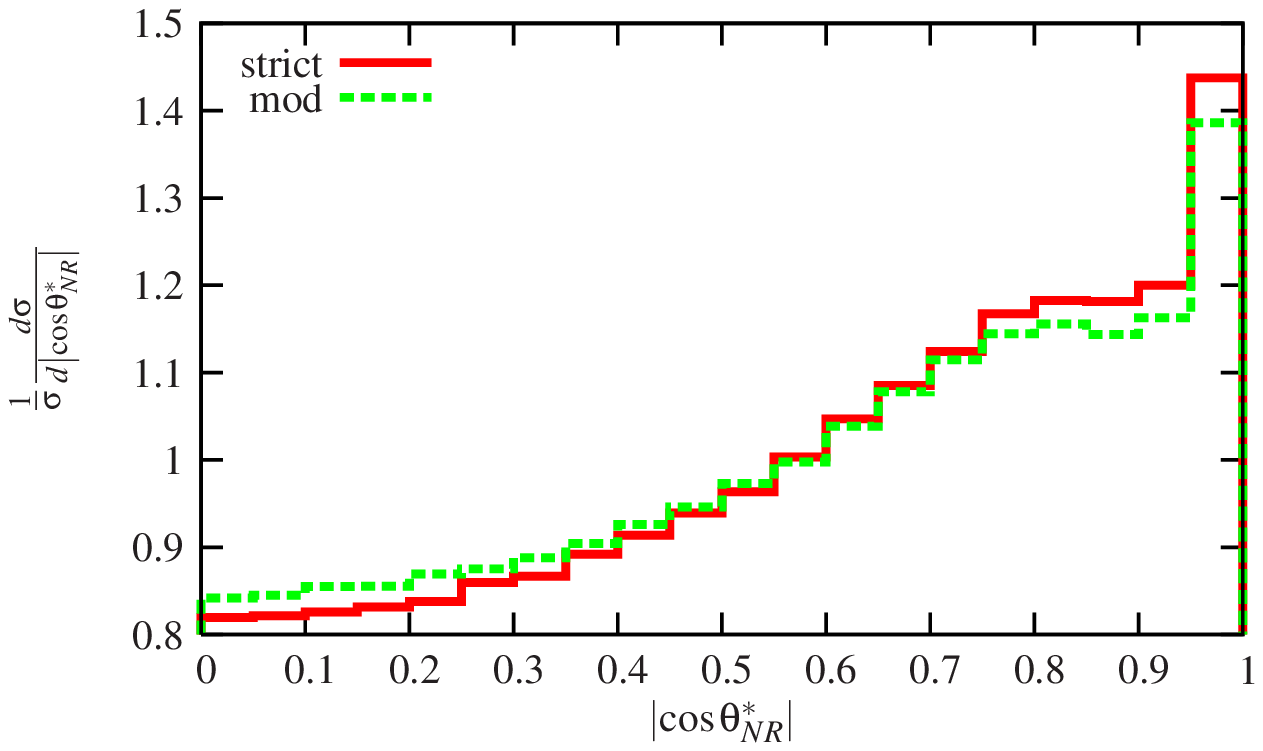}
&
\includegraphics[bb= 120 460 500 730,width=0.5\textwidth]{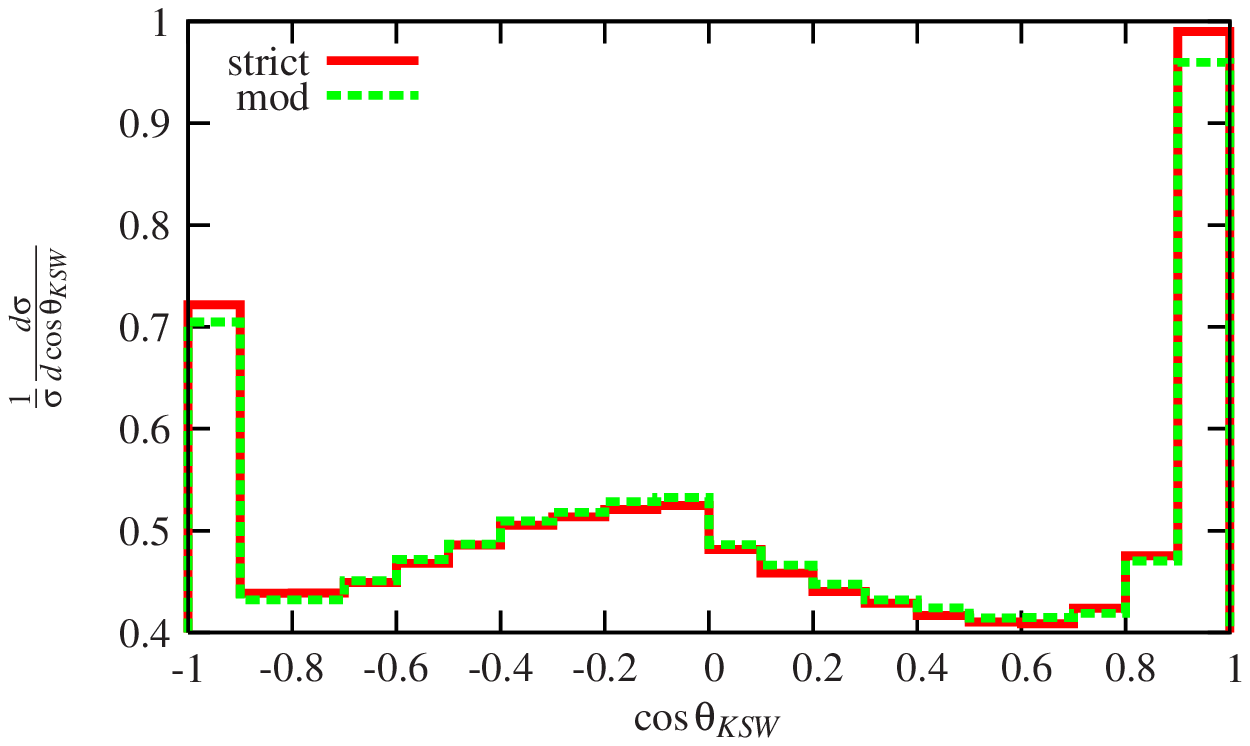}
\\
\includegraphics[bb= 120 460 500 730,width=0.5\textwidth]{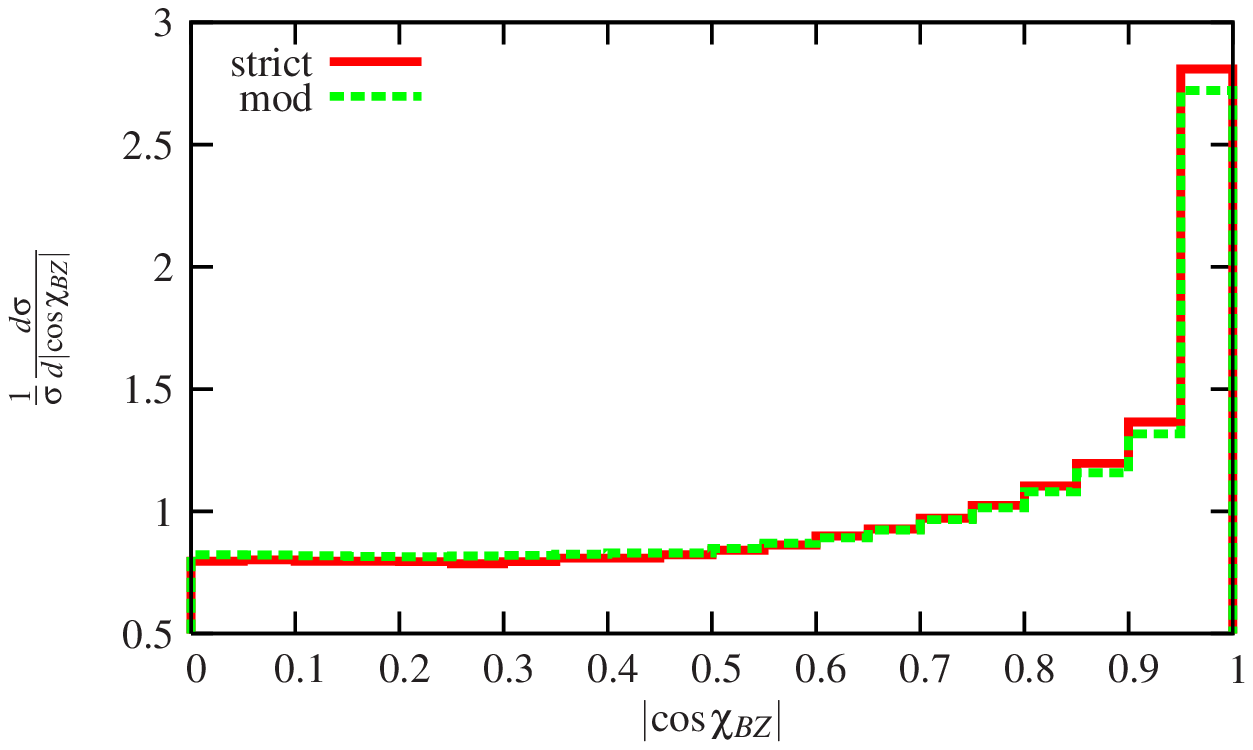}
&
\includegraphics[bb= 120 460 500 730,width=0.5\textwidth]{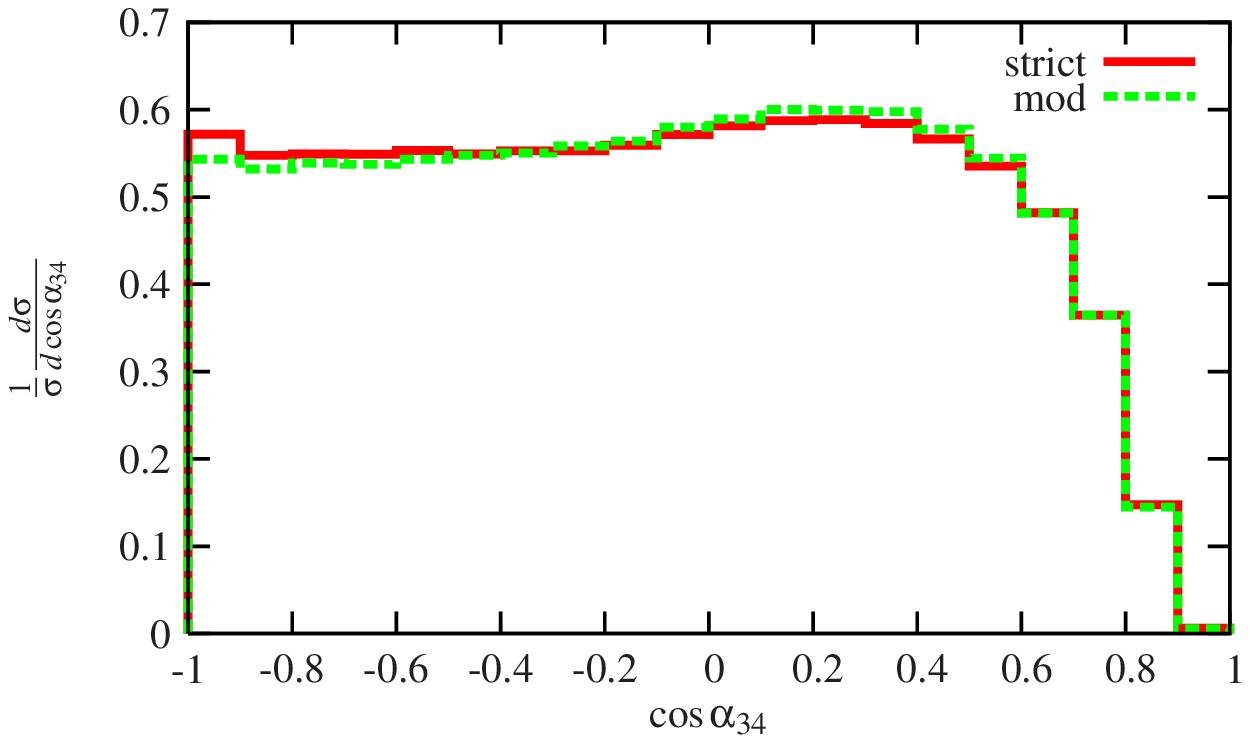}
\\
\end{tabular}
\end{center}
\caption{The distributions for the four-jet angles. From top-left to bottom-right:
The modified Nachtmann-Reiter angle,
the K\"orner-Schierholz-Willrodt angle,
the Bengtsson-Zerwas angle and
the angle $\alpha_{34}$ between the smallest energy jets. As cut-off parameter $Q_{min}=1\;\mbox{GeV}$ is used.
Shown are the result from the 
``strict leading-colour approximation'' and the ``modified leading-colour approximation''.}
\label{fig_fourjetangles}
\end{figure}

For electron-positron annihilation we use $\alpha_s(m_Z) = 0.118$ corresponding to
$\Lambda_5 = 88 \;\mbox{MeV}$.
We start the shower from the $2\rightarrow 2$ hard matrix element $e^+ e^- \rightarrow q \bar{q}$.
We first study the event shape variables thrust, the C-parameter and the D-parameter.
The distributions of the first moments of these observables are shown in figure~\ref{fig_thrust}
for two choices of the cut-off parameter: $Q_{min} = 1 \;\mbox{GeV}$ and $Q_{min} = 2 \;\mbox{GeV}$.
The distributions are normalised to unity.
The different prescriptions for the colour-treatment do not change the distributions significantly.
In figure~\ref{fig_fourjetangles} we show the distributions for the 
four-jet angles.
Again we start from the $2\rightarrow2$ hard matrix element.
The particles in an event are first clustered into jets,
defined according to the Durham algorithm \cite{Stirling:1991ds} with $y_{cut}=0.008$
and the $E$-scheme for the recombination.
Then events with exactly four jets are selected.
We consider 
the modified Nachtmann-Reiter angle \cite{Nachtmann:1982xr},
the K\"orner-Schierholz-Willrodt angle \cite{Korner:1980pv},
the Bengtsson-Zerwas angle \cite{Bengtsson:1988qg} and
the angle $\alpha_{34}$ between the jets with the smallest energy \cite{Abreu:1990ce}.
In the plots we show the results from the different options for the colour treatment for $Q_{min}=1 \;\mbox{GeV}$.
A variation of the cut-off scale does not change the distributions significantly.

\subsection{Hadron colliders}
\label{sect:results_hadron}

\begin{figure}[ht]
\begin{center}
\begin{tabular}{ll}
\includegraphics[bb= 120 460 500 730,width=0.5\textwidth]{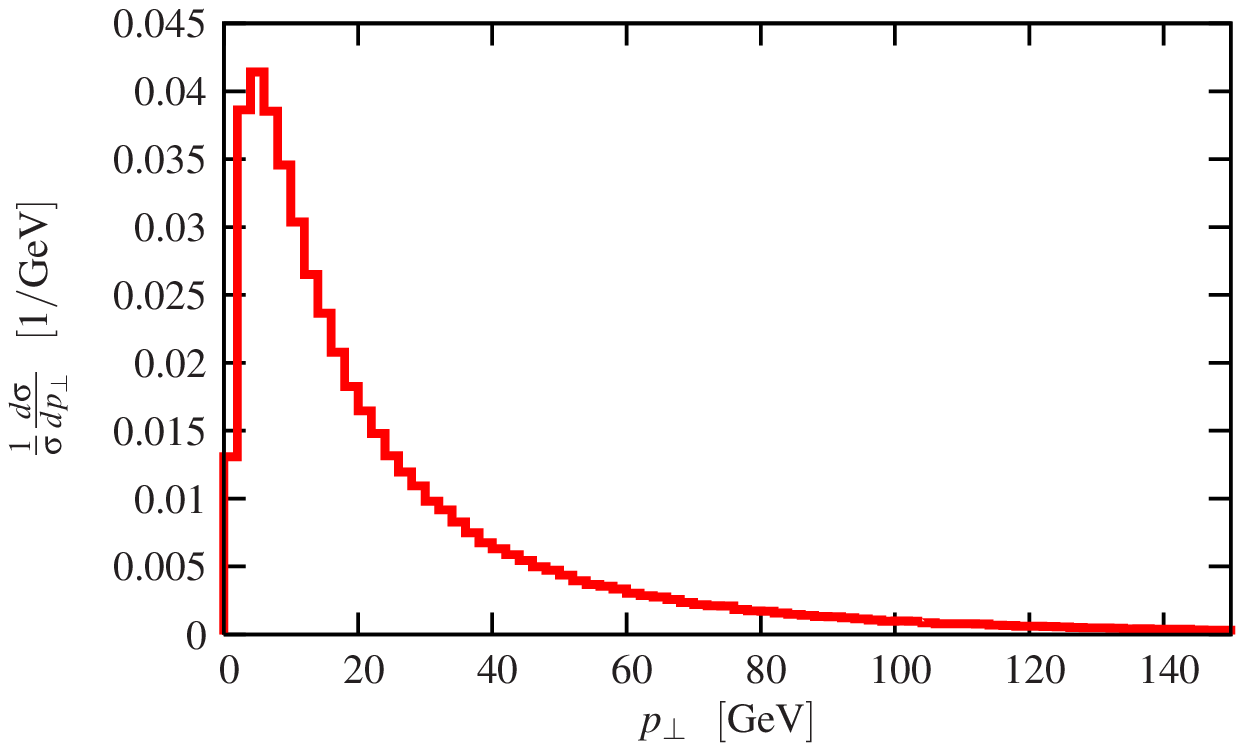}
&
\includegraphics[bb= 120 460 500 730,width=0.5\textwidth]{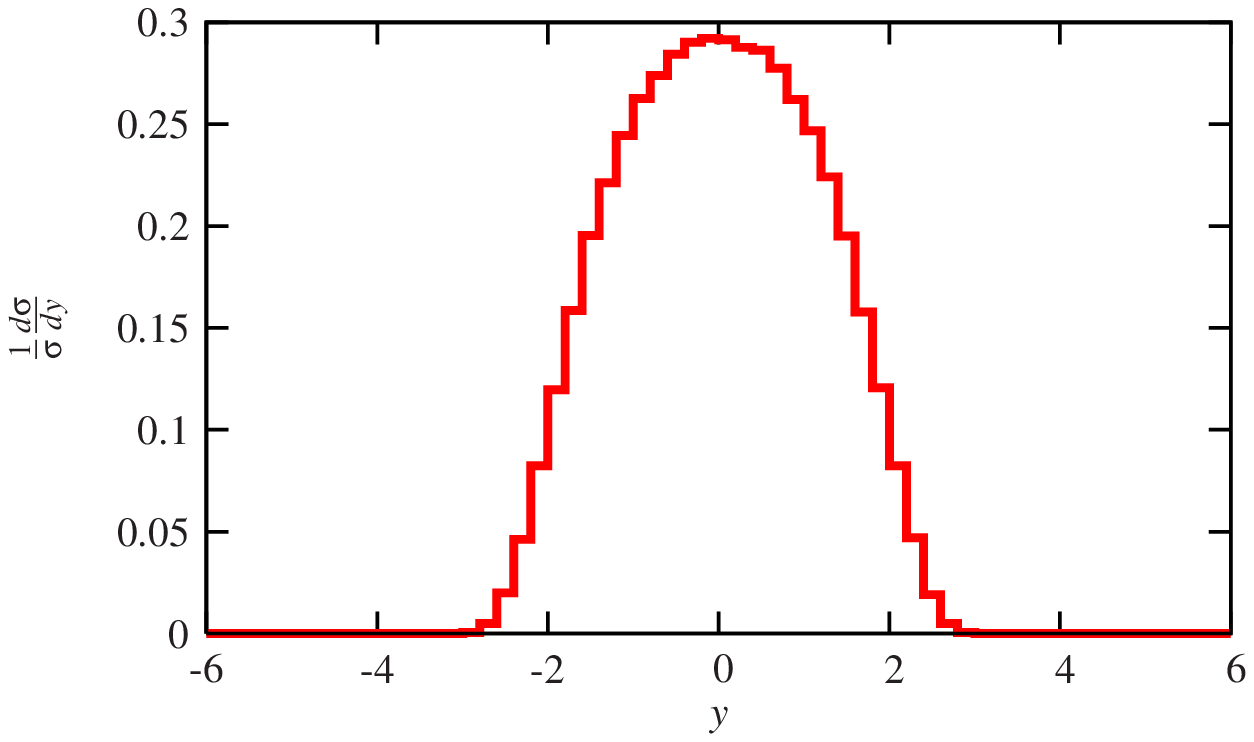}
\\
\includegraphics[bb= 120 460 500 730,width=0.5\textwidth]{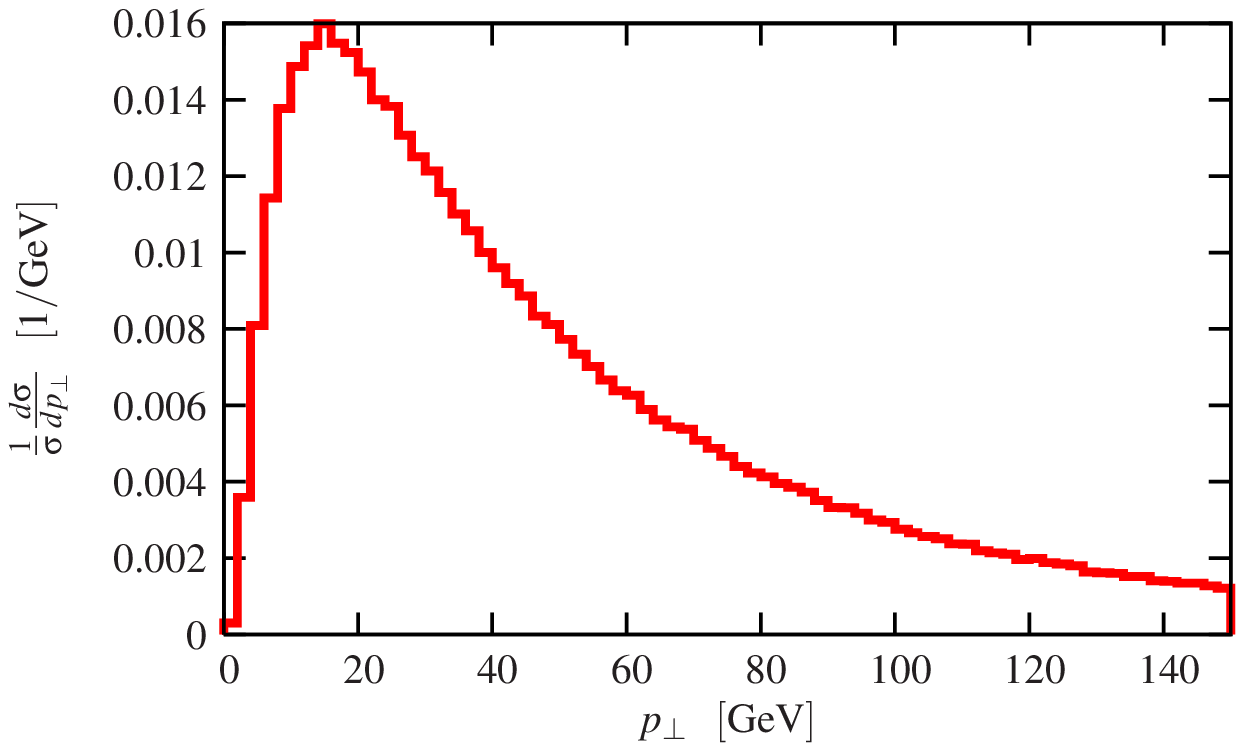}
&
\includegraphics[bb= 120 460 500 730,width=0.5\textwidth]{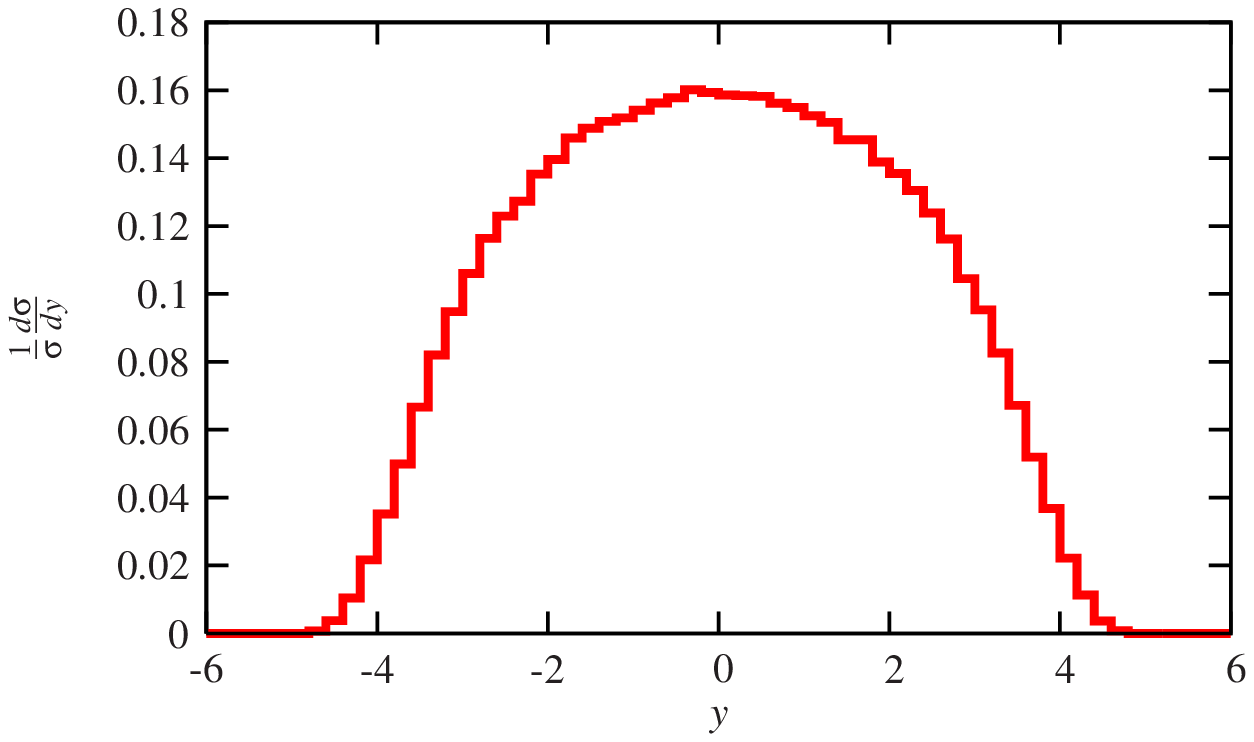}
\\
\end{tabular}
\end{center}
\caption{The transverse momentum distribution and the rapidity distribution of the lepton pair 
for $Z/\gamma^\ast$-production for the Tevatron and the LHC.
As cut-off parameter $Q_{min}=1\;\mbox{GeV}$ is used.}
\label{fig_Tevatron}
\end{figure}

For the Tevatron and the LHC we study $Z/\gamma^\ast$-production. We start from the 
$2 \rightarrow 2$ hard matrix element
$ q \bar{q} \rightarrow Z/\gamma^\ast \rightarrow l^+ l^-$.
As parton distribution functions we use the CTEQ 6L1 set \cite{Pumplin:2002vw,Stump:2003yu}.
For consistency we use here $\alpha_s(m_Z) = 0.130$ corresponding to
$\Lambda_5 = 165 \;\mbox{MeV}$.
The centre-of-mass energy we set to $\sqrt{s} = 1.96 \;\mbox{TeV}$ for the Tevatron and
to $\sqrt{s} = 14 \; \mbox{TeV}$ for the LHC.
We require a cut on the invariant mass of the lepton pair of
\bq
 m_{l^+ l^-} & > & 80 \; \mbox{GeV}.
\eq
As cut-off parameter for the parton shower we use $Q_{min} = 1 \; \mbox{GeV}$.
In figure~\ref{fig_Tevatron} we show the transverse momentum distribution and the 
rapidity distribution of the lepton pair for the Tevatron and the LHC.

% ------------------------------------------------------------------

\section{Summary}
\label{sect:summary}

In this paper we presented an implementation of a shower algorithm based 
on the dipole formalism. 
The formalism treats initial- and final-state partons on the same footing. 
The shower can be used for hadron colliders and electron-positron colliders.
We also included in the shower algorithm massive partons in the final state.
We studied numerical results for electron-positron annihilation, the Tevatron and the LHC.

\subsection*{Acknowledgments}

We would like to thank Zoltan Nagy for discussions and useful comments on the manuscript.

% ------------------------------------------------------------------

\begin{appendix}

\section{Sudakov factors for massless final-state partons}
\label{appendix:final_final}

In this appendix we discuss in more detail the Sudakov factors 
for massless final-state partons.
This case is simple enough that one integration can be done analytically.
The spin-averaged dipole subtraction terms in four dimensions are
\bq
{\cal P}_{q \rightarrow q g} & = &
   C_F
   \frac{8\pi \alpha_s(\mu^2)}{s_{ijk}} \frac{1}{y}
   \left[ \frac{2}{1-z(1-y)} - (1+z) \right],
 \nonumber \\
{\cal P}_{g \rightarrow g g} & = & 
   C_A
   \frac{8\pi \alpha_s(\mu^2)}{s_{ijk}} \frac{1}{y}
   \left[ \frac{2}{1-z(1-y)} + \frac{2}{1-(1-z)(1-y)} -4 + 2z(1-z) \right],
 \nonumber \\
{\cal P}_{g \rightarrow q \bar{q}} & = & 
   T_R 
   \frac{8\pi \alpha_s(\mu^2)}{s_{ijk}} \frac{1}{y}
   \left[ 1 - 2 z (1-z) \right],
\eq
with
\bq
 s_{ijk} = \left( p_i + p_j + p_k \right)^2 = (p_{\tilde{i}}+p_{\tilde{k}})^2.
\eq
The dipole phase space measure is
\bq
 \int d\phi_{unres} & = & 
  \frac{s_{ijk}}{16\pi^2} \int\limits_0^1 d\kappa \int\limits_{z_-(\kappa)}^{z_+(\kappa)} dz
     \frac{1}{4z(1-z)} \left( 1 - \frac{\kappa}{4 z (1-z)} \right),
\eq
with
\bq 
 z_{\pm}(\kappa) & = & \frac{1}{2} \left( 1 \pm \sqrt{1-\kappa} \right).
\eq
The strong coupling is evaluated at the scale $\mu^2=-k_\perp^2$:
\bq
 \alpha_s(\mu^2) & = & \alpha_s\left( \frac{1}{4} \kappa s_{ijk} \right).
\eq
The Sudakov factor is given by
\bq
 \Delta_{ij,k}(t_1,t_2) & = &
 \exp\left( - \int\limits_{t_2}^{t_1} dt {\cal C}_{\tilde{i},\tilde{k}} 
              \int d\phi_{unres} \delta\left(t-T_{\tilde{i},\tilde{k}} \right) {\cal P}_{ij,k} \right),
\eq
For the splitting $q\rightarrow q g$ we obtain
\bq
 \Delta_{ij,k}(t_1,t_2) & = &
  \exp\left\{ - {\cal C}_{\tilde{i},\tilde{k}} C_F 
   \int\limits_{\kappa_-}^{\kappa_+} \frac{d\kappa}{\kappa} 
   \frac{\alpha_s(\mu^2)}{2\pi} 
   \int\limits_{z_-(\kappa)}^{z_+(\kappa)} dz \; (1-y)
   \left[ \frac{2}{1-z(1-y)} - (1+z) \right]
  \right\},
\eq
with
\bq
 \kappa_- = 4 \frac{Q^2}{s_{ijk}} e^{t_2},
 \;\;\;
 \kappa_+ = \mbox{min}\left(1, 4 \frac{Q^2}{s_{ijk}} e^{t_1} \right),
 \;\;\;
 y = \frac{\kappa}{4 z (1-z)},
 \;\;\;
 \mu^2 = \frac{1}{4} \kappa s_{ijk}.
\eq
The integration over $z$ can be done analytically:
\bq
\lefteqn{
   \int dz \; (1-y) 
   \left[ \frac{2}{1-z(1-y)} - (1+z) \right]
 = 
 -\frac{1}{2} z^2 - z 
 + \frac{\kappa}{4} \left[ \ln z - 2 \ln(1-z) \right]
} & &
 \nonumber \\
 & &
 - \frac{4}{4 + \kappa} 
    \left[ \frac{1}{2} \kappa \ln z + \ln\left( \kappa + 4(1-z)^2 \right)
          + \sqrt{\kappa} \arctan\left( \frac{2}{\sqrt{\kappa}} (1-z) \right)
    \right].
\eq
The same holds for the other splittings.
Therefore we obtain for the Sudakov factors
\bq
 \Delta_{ij,k}(t_1,t_2) & = &
  \exp\left\{ - {\cal C}_{\tilde{i},\tilde{k}} C 
   \int\limits_{\kappa_-}^{\kappa_+} \frac{d\kappa}{\kappa} 
   \frac{\alpha_s\left(\frac{1}{4} \kappa s_{ijk} \right)}{2\pi} 
   \left(
   {\cal V}_{ij,k}\left(\kappa, z_+ \right)
   -
   {\cal V}_{ij,k}\left(\kappa, z_- \right)
   \right)
   \right\},
\eq
where $C$ is a colour factor and equal to
\bq
 C & = & \left\{
 \begin{array}{lll}
  C_F & \mbox{for} & q \rightarrow q g, \\
  C_A & \mbox{for} & g \rightarrow g g, \\
  T_R & \mbox{for} & g \rightarrow q \bar{q}. \\
 \end{array}
 \right.
\eq
The functions ${\cal V}_{ij,k}\left(\kappa, z \right)$ are given by
\bq
 {\cal V}_{qg,k}\left(\kappa, z \right)
 & = &
 -\frac{1}{2} z^2 - z 
 + \frac{\kappa}{4} \left[ \ln z - 2 \ln(1-z) \right]
 \nonumber \\
 &  &
 - \frac{4}{4 + \kappa} 
    \left[ \frac{1}{2} \kappa \ln z + \ln\left( \kappa + 4(1-z)^2 \right)
          - \sqrt{\kappa} \arctan\left( \frac{2}{\sqrt{\kappa}} (1-z) \right)
    \right],
 \nonumber \\
 {\cal V}_{gg,k}\left(\kappa, z \right)
 & = &
 - \frac{2}{3} z^3 + z^2 - 4 z 
 - \frac{1}{2} \kappa z 
 + \kappa \ln \frac{z}{1-z}
 + \frac{4}{4+\kappa} 
   \left[ \frac{1}{2} \kappa \ln \frac{1-z}{z}
 \right.
 \nonumber \\
 & &
 \left.
         + \ln \frac{\kappa + 4 z^2}{\kappa + 4(1-z)^2}
         - \sqrt{\kappa} \arctan\left( \frac{2 z}{\sqrt{\kappa}} \right)
         + \sqrt{\kappa} \arctan\left( \frac{2 (1-z)}{\sqrt{\kappa}} \right)
   \right],
 \nonumber \\
 {\cal V}_{gq,k}\left(\kappa, z \right)
 & = &
 \frac{2}{3} z^3 - z^2 + z + \frac{\kappa}{2} z 
 - \frac{\kappa}{4} \ln \frac{z}{1-z}.
\eq

% ------------------------------------------------------------------

\section{Insertion of emitted particles}
\label{sect:insertion}

In this appendix we list the relevant formul\ae\ 
for the insertion of one additional four-vector into a set of 
$n$ four-vectors.
This insertion satisfies momentum conservation and can be considered as the inverse of the
$(n+1) \rightarrow n$ phase space mapping of Catani and Seymour.
These insertion mappings are also useful for an efficient phase-space integration of the 
real emission contribution in NLO calculations.
Therefore we quote in addition the relevant phase space weights.
For the shower algorithm, these weights are not needed, 
as they are taken into account through the generation of the
shower.

\subsection{Insertion for final-state particles}

\subsubsection*{The massless case}

We start with the simplest case, where both the emitter and the spectator are in the final state and
all particles involved in the dipole splitting are massless.
The insertion procedure is identical to the one used in \cite{Weinzierl:1999yf}.
Given the four-vectors $\tilde{p}_{ij}$ and $\tilde{p}_k$ together with the three variables
$y$, $z$ and $\phi_s$ we would like to construct $p_i$, $p_j$ and $p_k$, such that
\bq
 p_i+p_j+p_k = \tilde{p}_{ij}+\tilde{p}_k,
 \;\;\; p_i^2 = p_j^2 = p_k^2 = 0.
\eq
In four dimensions we have for the phase space measure
\bq
 d\phi_{unres}
 & = & \frac{s_{ijk}}{32 \pi^3} 
       \int\limits_0^1 dy \; \left(1-y\right)
       \int\limits_0^1 dz \; 
       \int\limits_0^{2\pi} d\phi_s,
\eq
where $s_{ijk}=(\tilde{p}_{ij}+\tilde{p}_k)^2=(p_i+p_j+p_k)^2$.
It is convenient to work in the rest frame of $P=\tilde{p}_{ij}+\tilde{p}_k=p_i+p_j+p_k$.
We shall orient the frame in such a way, that the spatial components of $\tilde{p}_k$ are along the $z$-direction.
When used as a phase space generator we set
\bq
 y = u_1, \;\;\; z = u_2 \;\;\; \phi_s = 2 \pi u_3,
\eq
where $u_1$, $u_2$ and $u_3$ are three uniformly distributed random numbers in $[0,1]$.
From
\bq
 y = \frac{s_{ij}}{s_{ij}+s_{ik}+s_{jk}},
 & & 
 z = \frac{s_{ik}}{s_{ik} + s_{jk}}
\eq
we obtain
\bq
 s_{ij} = y P^2,
 \;\;\;
 s_{ik} = z (1-y) P^2, 
 \;\;\;
 s_{jk} & = & (1-z) (1-y) P^2.
\eq
If $s_{ij} < s_{jk}$ we want to have $p_k' \rightarrow p_k$ as $s_{ij} \rightarrow 0$.
Define
\bq
E_i = \frac{s_{ij}+s_{ik}}{2 \sqrt{s_{ijk}}}, \hspace{1cm}
E_j = \frac{s_{ij}+s_{jk}}{2 \sqrt{s_{ijk}}} ,\hspace{1cm}
E_k = \frac{s_{ik}+s_{jk}}{2 \sqrt{s_{ijk}}}, 
\eq
\bq
\theta_{ik}  =  \arccos \left( 1 - \frac{s_{ik}}{2 E_i E_k} \right), & &
\theta_{jk}  =  \arccos \left( 1 - \frac{s_{jk}}{2 E_j E_k} \right) .
\eq
In our coordinate system we have
\bq
p_i' & = & E_i ( 1, \sin \theta_{ik} \cos(\phi_s+\pi), \sin \theta_{ik} \sin(\phi_s+\pi), \cos \theta_{ik} ) ,\nonumber \\
p_j' & = & E_j ( 1, \sin \theta_{jk} \cos \phi_s, \sin \theta_{jk} \sin \phi_s, \cos \theta_{jk} ) ,\nonumber \\
p_k' & = & E_k ( 1, 0, 0, 1) .
\eq
The momenta $p_i'$, $p_j'$ and $p_k'$ are related to the momenta $p_i$, $p_j$ and $p_k$ by a sequence of
Lorentz transformations back to the original frame
\bq
\label{lorentzchain}
p_i & = & \Lambda_{boost} \Lambda_{xy}(\phi) \Lambda_{xz}(\theta) p_i'
\eq
and analogously for the other two momenta. 
The explicit formul\ae\ for the Lorentz transformations are obtained as follows :
Let $|P| = \sqrt{(\tilde{p}_{ij}+\tilde{p}_k)^2}$ and denote by $\hat{p}_k$ the coordinates of the hard momentum 
$\tilde{p}_k$ in the centre of
mass system of $\tilde{p}_{ij}+\tilde{p}_k$. $\hat{p}_k$ is given by
\bq
 \hat{p}_k & = & 
  \left( 
        \frac{E_P}{|P|} \tilde{E}_k - \frac{\vec{\tilde{p}}_k \cdot \vec{P}}{|P|}, 
        \vec{\tilde{p}}_k + \left( \frac{\vec{\tilde{p}}_k \cdot \vec{P}}{|P| (E_P+|P|)} 
        - \frac{\tilde{E}_k}{|P|} \right) \vec{P}
  \right)
\eq
The angles are then given by
\bq
\theta = \arccos \left( \frac{2 \hat{E}_k E_k' - 2 \hat{p}_k \cdot p_k'}{2 \left| \hat{\vec{p}}_k \right| \left| \vec{p}_k' \right|} \right),
 & &
\phi = \arctan\left( \frac{\hat{p}_k^y}{\hat{p}_k^x} \right).
\eq
For the case considered here particle $k$ is massless and the formula for $\theta$ reduces to
\bq
\theta = \arccos \left( 1 - \frac{2 \hat{p}_k \cdot p_k'}{2 \hat{p}_k^t p_k^{t'}} \right).
\eq
The explicit form of the rotations is
\bq
\Lambda_{xz}(\theta) =  
\left(
\begin{array}{cccc}
1 & 0 & 0 & 0 \\
0 & \cos \theta & 0 & \sin \theta \\
0 & 0 & 1 & 0 \\
0 & - \sin \theta & 0 & \cos \theta \\
\end{array}
\right), 
& &
\Lambda_{xy} (\phi) = 
\left(
\begin{array}{cccc}
1 & 0 & 0 & 0 \\
0 & \cos \phi & - \sin \phi & 0 \\
0 & \sin \phi & \cos \phi & 0 \\
0 & 0 & 0 & 1 \\
\end{array}
\right).
\eq
The boost $p = \Lambda_{boost} q $ is given by 
\bq
p & = & 
 \left( 
       \frac{E_P}{|P|} E_q + \frac{\vec{q} \cdot \vec{P}}{|P|}, 
       \vec{q}+ \left( \frac{\vec{q} \cdot \vec{P}}{|P| (E_P+|P|)} 
       + \frac{E_q}{|P|} \right) \vec{P}
 \right).
\eq
The weight is given by
\bq
 w & = & \frac{s_{ijk}}{16 \pi^2} \left( 1-y \right).
\eq

\subsubsection*{The massive case}

We now consider the case of final state particles with arbitrary masses:
\bq
 \tilde{p}_{ij}^2 = m_{ij}^2,
 \;\;\;
 p_i^2 = m_i^2,
 \;\;\;
 p_j^2 = m_j^2,
 \;\;\;
 \tilde{p}_k^2 = p_k^2 = m_k^2.
\eq
The dipole phase space reads \cite{Catani:2002hc}
\bq
 d\phi_{unres}
 & = & \frac{s_{ijk}}{32 \pi^3} 
       \left( 1 - \mu_i^2 - \mu_j^2 - \mu_k^2 \right)^2
       \left[ \lambda\left(1,\mu_{ij}^2,\mu_k^2\right) \right]^{- \frac{1}{2}}
       \int\limits_{y_-}^{y_+} dy \; \left(1-y\right)
       \int\limits_ {z_-}^{z_+} dz \; 
       \int\limits_0^{2\pi} d\phi_s,
\eq
where
\bq
\label{def_massive_1}
 s_{ijk} = \left( \tilde{p}_{ij} + \tilde{p}_k \right)^2,
 \;\;\;
 \mu_l = \frac{m_l}{\sqrt{s_{ijk}}},
 \;\;\;
 \lambda(x,y,z) & = & x^2 + y^2 + z^2 - 2 xy - 2 yz - 2 zy.
\eq
The integration boundaries are given by
\bq
\label{def_massive_2}
 y_+ = 1 - \frac{2\mu_k(1-\mu_k)}{1-\mu_i^2-\mu_j^2-\mu_k^2},
 & &
 y_- = \frac{2\mu_i\mu_j}{1-\mu_i^2-\mu_j^2-\mu_k^2}.
\eq
\bq
\label{def_massive_3}
 z_\pm & = & \frac{2\mu_i^2 + \left( 1-\mu_i^2-\mu_j^2-\mu_k^2 \right) y}
                  {2\left[ \mu_i^2+\mu_j^2+\left(1-\mu_i^2-\mu_j^2-\mu_k^2\right)y\right]}
             \left( 1 \pm v_{ij,i} v_{ij,k} \right).
\eq
The general formula for the relative velocities is
$ v_{p,q} =  \sqrt{1 - p^2 q^2/(p q)}$.
In our case the relative velocities are given by
\bq
\label{def_massive_4}
 v_{ij,k} & = & \frac{\sqrt{\left[2\mu_k^2+\left(1-\mu_i^2-\mu_j^2-\mu_k^2\right)\left(1-y\right)\right]^2-4\mu_k^2}}
                     {\left(1-\mu_i^2-\mu_j^2-\mu_k^2\right)\left(1-y\right)},
 \nonumber \\
 v_{ij,i} & = & \frac{\sqrt{\left(1-\mu_i^2-\mu_j^2-\mu_k^2\right)^2 y^2-4\mu_i^2\mu_j^2}}
                     {\left(1-\mu_i^2-\mu_j^2-\mu_k^2\right)y+2\mu_i^2}.
\eq
For the phase space generation we set
\bq
 y = \left( y_+ - y_- \right) u_1 + y_-,
 \;\;\;
 z = \left( z_+ - z_- \right) u_2 + z_-,
 \;\;\,
 \phi_s = 2 \pi u_3.
\eq
We again work in the rest frame of $P=\tilde{p}_{ij}+\tilde{p}_k=p_i+p_j+p_k$,
such that the spatial components of $\tilde{p}_k$ are along the $z$-direction:
\bq
 \tilde{p}_{ij} = \left( \tilde{E}_{ij}, 0,0, -\left| \vec{\tilde{p}}_k \right| \right),
 & &
 \tilde{p}_{k} = \left( \tilde{E}_k, 0,0, \left| \vec{\tilde{p}}_k \right| \right).
\eq
For the invariants we have
\bq
 2 p_i p_j & = & y \left( P^2 -m_i^2 -m_j^2 -m_k^2 \right),
 \nonumber \\
 2 p_i p_k & = & z \left(1-y\right) \left( P^2 -m_i^2 -m_j^2 -m_k^2 \right),
 \nonumber \\
 2 p_j p_k & = & \left(1-z\right) \left(1-y\right) \left( P^2 -m_i^2 -m_j^2 -m_k^2 \right).
\eq
The invariants are related to $y$ and $z$ as follows:
\bq
 y = \frac{2 p_i p_j}{2 p_i p_j + 2 p_i p_k + 2 p_j p_k},
 & & 
 z = \frac{2 p_i p_k}{2 p_i p_k + 2 p_j p_k}.
\eq
In our chosen frame
\bq
p_i' & = & \left|\vec{p}_i\right| ( \frac{E_i}{\left|\vec{p}_i\right|}, \sin \theta_{ik} \cos(\phi_s+\pi), \sin \theta_{ik} \sin(\phi_s+\pi), \cos \theta_{ik} ) ,\nonumber \\
p_j' & = & \left|\vec{p}_j\right| ( \frac{E_j}{\left|\vec{p}_j\right|}, \sin \theta_{jk} \cos \phi_s, \sin \theta_{jk} \sin \phi_s, \cos \theta_{jk} ) ,\nonumber \\
p_k' & = & \left|\vec{p}_k\right| ( \frac{E_k}{\left|\vec{p}_k\right|}, 0, 0, 1) .
\eq
The energies are obtained from the invariants as follows:
\bq
E_i & = & \frac{s_{ijk}-2p_jp_k + m_i^2 - m_j^2 - m_k^2}{2\sqrt{s_{ijk}}},
 \nonumber \\
E_j & = & \frac{s_{ijk}-2p_ip_k - m_i^2 + m_j^2 - m_k^2}{2\sqrt{s_{ijk}}},
 \nonumber \\
E_k & = & \frac{s_{ijk}-2p_ip_j - m_i^2 - m_j^2 + m_k^2}{2\sqrt{s_{ijk}}}.
\eq
For the angles we have
\bq
 \theta_{ik} = \arccos\left( \frac{2 E_i E_k - 2 p_i p_k}{2 \left|\vec{p}_i\right| \left|\vec{p}_k\right|} \right),
 & &
 \theta_{jk} = \arccos\left( \frac{2 E_j E_k - 2 p_j p_k}{2 \left|\vec{p}_j\right| \left|\vec{p}_k\right|} \right).
\eq
The momenta $p_i'$, $p_j'$ and $p_k'$ are related to the momenta $p_i$, $p_j$ and $p_k$ 
by the same sequence of Lorentz transformations as in eq.~(\ref{lorentzchain}).
The weight is
\bq
 w & = &
  \frac{s_{ijk}}{16 \pi^2} 
       \left( 1 - \mu_i^2 - \mu_j^2 - \mu_k^2 \right)^2
       \left[ \lambda\left(1,\mu_{ij}^2,\mu_k^2\right) \right]^{- \frac{1}{2}}
       \left(1-y\right) \left( y_+ - y_- \right) \left( z_+ - z_- \right).
\eq

% ----------------------------

\subsection{Insertion for an antenna between an initial-state and a final state}

\subsubsection*{The massless case}

Here the $(n+1)$-particle phase space is given by a convolution:
\bq
 d\phi_{n+1}
 & = &
 \int\limits_0^1 dx \; d\phi_n(x p_a) \; d\phi_{dipole}.
\eq
The dipole phase space reads:
\bq
 d\phi_{dipole}
 & = & 
 \frac{\left| 2 \tilde{p}_{ij}p_a \right|}{32 \pi^3} \int\limits_0^1 dz \int\limits_0^{2\pi} d\phi_s.
\eq
The angle
$\phi_s$ parametrises the solid angle perpendicular to $\tilde{p}_{ij}$ and $x p_a$.
Therefore we can treat the case of a final-state emitter with an initial-state spectator 
as well as the case of an initial-state emitter with a final-state spectator at the same time.
$x$ and $z$ are related to the invariants as follows:
\bq
 x = \frac{-2p_ip_a -2p_jp_a - 2 p_i p_j}{-2p_ip_a-2p_jp_a},
 & &
 z = \frac{-2p_ip_a}{-2p_ip_a - 2p_jp_a}.
\eq
For the phase space generation we set
\bq
 x = 1-u_1,
 \;\;\;
 z = u_2,
 \;\;\,
 \phi_s = 2 \pi u_3.
\eq
We denote $Q=\tilde{p}_{ij}+xp_a = p_i+p_j+p_a$. It is convenient to work in the rest frame of
$P=p_i+p_j=Q-p_a$ and to orient the frame such that $p_a$ is along the $z$-axis.
For the invariants we have
\bq
 2p_ip_j 
 = \left( -Q^2 \right) \frac{1-x}{x},
 \;\;\;
 2p_ip_a = \frac{z}{x} Q^2,
 \;\;\;
 2p_jp_a = \frac{1-z}{x} Q^2.
\eq
In this frame
\bq
p_i' & = & E_i ( 1, \sin \theta_{ia} \cos \phi_s, \sin \theta_{ia} \sin \phi_s, \cos \theta_{ia} ) ,\nonumber \\
p_j' & = & E_i ( 1, -\sin \theta_{ia} \cos \phi_s, -\sin \theta_{ia} \sin \phi_s, -\cos \theta_{ia} ) ,\nonumber \\
p_a' & = & ( -\left|E_a\right|, 0, 0, \left|E_a\right| \mbox{sign}({p_a^z}')) .
\eq
We have
\bq
 E_i = \frac{1}{2} \left| P \right|,
 &  &
 E_a = \frac{1}{\left|P\right|} \left( P \cdot p_a \right),
 \;\;\;
 \theta_{ia} = \arccos\left[ \mbox{sign}({p_a^z}') \left( -1 + \frac{2p_ip_a}{2E_iE_a} \right) \right].
\eq
The momenta $p_i'$, $p_j'$ are again related to the momenta $p_i$, $p_j$
by a sequence of Lorentz transformations as in eq.~(\ref{lorentzchain}).
The weight is given by
\bq
 w & = & \frac{\left|Q^2\right|}{16 \pi^2 x}.
\eq

\subsubsection*{The massive case}

The dipole phase space now reads:
\bq
 d\phi_{dipole}
 & = & 
 \frac{\left| 2 \tilde{p}_{ij}p_a \right|}{32 \pi^3} \int\limits_{z_-}^{z_+} dz \int\limits_0^{2\pi} d\phi_s.
\eq
The integration boundaries are given by
\bq
 z_+ = 1,
 & &
 z_- = \frac{\mu^2}{1-x+\mu^2}.
\eq
where
\bq
 \mu^2 & = & \frac{m_i^2}{\left| 2 \tilde{p}_{ij}p_a \right|} = \frac{x m_i^2}{\left| Q^2-m_i^2 \right|}.
\eq
We consider only the case where $m_{\tilde{ij}}=m_i=m$ and all other masses are zero.
For the phase space generation we set
\bq
 x = 1-u_1,
 \;\;\;
 z = \left( z_+ - z_- \right) u_2 + z_-,
 \;\;\,
 \phi_s = 2 \pi u_3.
\eq
For the invariants we have now
\bq 
 2p_ip_j 
 = \left(-Q^2+m_i^2 \right) \frac{1-x}{x},
 \;\;\;
 2p_ip_a = \frac{z}{x} \left(Q^2-m_i^2\right),
 \;\;\;
 2p_jp_a = \frac{1-z}{x} \left(Q^2-m_i^2\right).
\eq
We parametrise the momenta as
\bq
p_i' & = & \left| \vec{p}_i \right|  ( \frac{E_i}{| \vec{p}_i |} , \sin \theta_{ia} \cos \phi_s, \sin \theta_{ia} \sin \phi_s, \cos \theta_{ia} ) ,\nonumber \\
p_j' & = & \left| \vec{p}_i \right| ( 1, -\sin \theta_{ia} \cos \phi_s, -\sin \theta_{ia} \sin \phi_s, -\cos \theta_{ia} ) ,\nonumber \\
p_a' & = & ( -\left|E_a\right|, 0, 0, \left|E_a\right| \mbox{sign}({p_a^z}')) .
\eq
Then
\bq
 E_i = \frac{P^2+m_i^2}{2 \left| P \right|},
 & &
 E_a = \frac{1}{\left|P\right|} \left( P \cdot p_a \right),
 \;\;\;
 \theta_{ia} = \arccos\left[ \mbox{sign}({p_a^z}') \frac{\left(2E_iE_a-2p_ip_a\right)}{2\left|\vec{p}_i\right|(-E_a)} \right].
\eq
The momenta $p_i'$, $p_j'$ are again related to the momenta $p_i$, $p_j$
by a sequence of Lorentz transformations as in eq.~(\ref{lorentzchain}).
The weight is given by
\bq
 w & = & \frac{\left|Q^2-m_i^2\right|}{16 \pi^2 x} \left( z_+ - z_- \right).
\eq

% --------------------------------------

\subsection{Insertion for an initial-state antenna}

Here we only have to consider the case where all particles are massless.
In this case we transform all the final state momenta.
The $(n+1)$-particle phase space is given by a convolution:
\bq
 d\phi_{n+1}
 & = &
 \int\limits_0^1 dx \; d\phi_n(x p_a) \; d\phi_{dipole}.
\eq
The dipole phase space reads:
\bq
 d\phi_{dipole}
 & = & 
 \frac{\left| 2 p_a p_b \right|}{32 \pi^3} \int\limits_0^{1-x} dv \int\limits_0^{2\pi} d\phi_s.
\eq
The variable $v$ is given by
\bq
 v & = & \frac{-2p_a p_i}{2p_a p_b}.
\eq
For the phase space generation we set
\bq
 x = 1-u_1,
 \;\;\;
 v = (1-x)(1-u_2),
 \;\;\,
 \phi_s = 2 \pi u_3.
\eq
We denote
\bq
 K = -p_a-p_b-p_i, 
 & &
 \tilde{K} = -\tilde{p}_{ai} - p_b.
\eq
We have
\bq
 p_a & = & \frac{1}{x} \tilde{p}_{ai},
 \nonumber \\
 p_i & = & \Lambda_{boost} E_i \left( 1, \sin \theta_{ia} \cos \phi_s, \sin \theta_{ia} \sin \phi_s, \cos \theta_{ia} \right)
 \nonumber \\
 p_b & = & p_b,
\eq
with $E_i$ and $\theta_{ia}$ given in the rest frame of $p_a+p_b$ by
\bq
 E_a = - \frac{1}{2} \sqrt{2p_a p_b},
 \;\;\;
 E_i = \frac{\tilde{K}^2-2 p_a p_b}{4 E_a},
 \;\;\;
 \theta_{ia} = \arccos\left[ \mbox{sign}(\hat{p}_a^z) \left(-1+\frac{2 p_i p_a}{2 E_i E_a} \right) \right].
\eq
$\hat{p}_a$ denotes $p_a$ in the rest frame of  $p_a+p_b$.
$\Lambda_{boost}$ transforms from the rest frame of $p_a+p_b$ to the lab frame.
All other final state momenta are transformed with
\bq
 \Lambda^{-1} & = & g^{\mu\nu} 
 - 2 \frac{\left(K+\tilde{K}\right)^\mu \left(K+\tilde{K}\right)^\nu}{\left(K+\tilde{K}\right)^2}
 + 2 \frac{K^\mu \tilde{K}^\nu}{K^2}.
\eq
The weight is given by
\bq
 w & = & \frac{\left|\tilde{K}^2\right|}{16 \pi^2 x} (1-x).
\eq

\end{appendix}

% references
\bibliography{/home/stefanw/notes/biblio}

\begin{thebibliography}{10}

\bibitem{Sjostrand:2006za}
T.~Sj{\"o}strand, S.~Mrenna, and P.~Skands,
\newblock JHEP {\bf 05}, 026 (2006), hep-ph/0603175.
%%CITATION = HEP-PH/0603175;%%

\bibitem{Bertini:2000uh}
M.~Bertini, L.~L{\"o}nnblad, and T.~Sj{\"o}strand,
\newblock Comput. Phys. Commun. {\bf 134}, 365 (2001), hep-ph/0006152.
%%CITATION = HEP-PH 0006152;%%

\bibitem{Corcella:2000bw}
G.~Corcella {\em et~al.},
\newblock JHEP {\bf 01}, 010 (2001), hep-ph/0011363.
%%CITATION = HEP-PH 0011363;%%

\bibitem{Gieseke:2003hm}
S.~Gieseke, A.~Ribon, M.~H. Seymour, P.~Stephens, and B.~Webber,
\newblock JHEP {\bf 02}, 005 (2004), hep-ph/0311208.
%%CITATION = HEP-PH 0311208;%%

\bibitem{Gleisberg:2003xi}
T.~Gleisberg {\em et~al.},
\newblock JHEP {\bf 02}, 056 (2004), hep-ph/0311263.
%%CITATION = HEP-PH 0311263;%%

\bibitem{Marchesini:1983bm}
G.~Marchesini and B.~R. Webber,
\newblock Nucl. Phys. {\bf B238}, 1 (1984).
%%CITATION = NUPHA,B238,1;%%

\bibitem{Webber:1983if}
B.~R. Webber,
\newblock Nucl. Phys. {\bf B238}, 492 (1984).
%%CITATION = NUPHA,B238,492;%%

\bibitem{Gustafson:1986db}
G.~Gustafson,
\newblock Phys. Lett. {\bf B175}, 453 (1986).
%%CITATION = PHLTA,B175,453;%%

\bibitem{Gustafson:1987rq}
G.~Gustafson and U.~Pettersson,
\newblock Nucl. Phys. {\bf B306}, 746 (1988).
%%CITATION = NUPHA,B306,746;%%

\bibitem{Andersson:1989ki}
B.~Andersson, G.~Gustafson, and L.~L{\"o}nnblad,
\newblock Nucl. Phys. {\bf B339}, 393 (1990).
%%CITATION = NUPHA,B339,393;%%

\bibitem{Andersson:1988gp}
B.~Andersson, G.~Gustafson, L.~L{\"o}nnblad, and U.~Pettersson,
\newblock Z. Phys. {\bf C43}, 625 (1989).
%%CITATION = ZEPYA,C43,625;%%

\bibitem{Lonnblad:1992tz}
L.~L{\"o}nnblad,
\newblock Comput. Phys. Commun. {\bf 71}, 15 (1992).
%%CITATION = CPHCB,71,15;%%

\bibitem{Kato:1986sg}
K.~Kato and T.~Munehisa,
\newblock Phys. Rev. {\bf D36}, 61 (1987).
%%CITATION = PHRVA,D36,61;%%

\bibitem{Kato:1988ii}
K.~Kato and T.~Munehisa,
\newblock Phys. Rev. {\bf D39}, 156 (1989).
%%CITATION = PHRVA,D39,156;%%

\bibitem{Kato:1990as}
K.~Kato and T.~Munehisa,
\newblock Comput. Phys. Commun. {\bf 64}, 67 (1991).
%%CITATION = CPHCB,64,67;%%

\bibitem{Tanaka:2005rm}
H.~Tanaka, T.~Sugiura, and Y.~Wakabayashi,
\newblock Prog. Theor. Phys. {\bf 114}, 477 (2005), hep-ph/0510185.
%%CITATION = HEP-PH/0510185;%%

\bibitem{Catani:2001cc}
S.~Catani, F.~Krauss, R.~Kuhn, and B.~R. Webber,
\newblock JHEP {\bf 11}, 063 (2001), hep-ph/0109231.
%%CITATION = HEP-PH 0109231;%%

\bibitem{Krauss:2002up}
F.~Krauss,
\newblock JHEP {\bf 08}, 015 (2002), hep-ph/0205283.
%%CITATION = HEP-PH 0205283;%%

\bibitem{Schalicke:2005nv}
A.~Sch{\"a}licke and F.~Krauss,
\newblock JHEP {\bf 07}, 018 (2005), hep-ph/0503281.
%%CITATION = HEP-PH/0503281;%%

\bibitem{Mangano:2001xp}
M.~L. Mangano, M.~Moretti, and R.~Pittau,
\newblock Nucl. Phys. {\bf B632}, 343 (2002), hep-ph/0108069.
%%CITATION = HEP-PH/0108069;%%

\bibitem{Mrenna:2003if}
S.~Mrenna and P.~Richardson,
\newblock JHEP {\bf 05}, 040 (2004), hep-ph/0312274.
%%CITATION = HEP-PH 0312274;%%

\bibitem{Baer:1991qf}
H.~Baer and M.~H. Reno,
\newblock Phys. Rev. {\bf D44}, 3375 (1991).
%%CITATION = PHRVA,D44,3375;%%

\bibitem{Baer:1992ca}
H.~Baer and M.~H. Reno,
\newblock Phys. Rev. {\bf D45}, 1503 (1992).
%%CITATION = PHRVA,D45,1503;%%

\bibitem{Friberg:1999fh}
C.~Friberg and T.~Sj{\"o}strand,
\newblock (1999), hep-ph/9906316.
%%CITATION = HEP-PH 9906316;%%

\bibitem{Mrenna:1999mq}
S.~Mrenna,
\newblock (1999), hep-ph/9902471.
%%CITATION = HEP-PH 9902471;%%

\bibitem{Potter:2000an}
B.~P{\"o}tter,
\newblock Phys. Rev. {\bf D63}, 114017 (2001), hep-ph/0007172.
%%CITATION = HEP-PH 0007172;%%

\bibitem{Potter:2001ej}
B.~P{\"o}tter and T.~Sch{\"o}rner,
\newblock Phys. Lett. {\bf B517}, 86 (2001), hep-ph/0104261.
%%CITATION = HEP-PH 0104261;%%

\bibitem{Dobbs:2001gb}
M.~Dobbs,
\newblock Phys. Rev. {\bf D64}, 034016 (2001), hep-ph/0103174.
%%CITATION = HEP-PH 0103174;%%

\bibitem{Dobbs:2001dq}
M.~Dobbs,
\newblock Phys. Rev. {\bf D65}, 094011 (2002), hep-ph/0111234.
%%CITATION = HEP-PH 0111234;%%

\bibitem{Collins:2001fm}
J.~Collins,
\newblock Phys. Rev. {\bf D65}, 094016 (2002), hep-ph/0110113.
%%CITATION = HEP-PH/0110113;%%

\bibitem{Frixione:2002ik}
S.~Frixione and B.~R. Webber,
\newblock JHEP {\bf 06}, 029 (2002), hep-ph/0204244.
%%CITATION = HEP-PH/0204244;%%

\bibitem{Kramer:2003jk}
M.~Kr{\"a}mer and D.~E. Soper,
\newblock Phys. Rev. {\bf D69}, 054019 (2004), hep-ph/0306222.
%%CITATION = HEP-PH 0306222;%%

\bibitem{Soper:2003ya}
D.~E. Soper,
\newblock Phys. Rev. {\bf D69}, 054020 (2004), hep-ph/0306268.
%%CITATION = HEP-PH 0306268;%%

\bibitem{Nason:2004rx}
P.~Nason,
\newblock JHEP {\bf 11}, 040 (2004), hep-ph/0409146.
%%CITATION = HEP-PH/0409146;%%

\bibitem{Nagy:2005aa}
Z.~Nagy and D.~E. Soper,
\newblock JHEP {\bf 10}, 024 (2005), hep-ph/0503053.
%%CITATION = HEP-PH/0503053;%%

\bibitem{Nagy:2006kb}
Z.~Nagy and D.~E. Soper,
\newblock (2006), hep-ph/0601021.
%%CITATION = HEP-PH/0601021;%%

\bibitem{Nagy:2007ty}
Z.~Nagy and D.~E. Soper,
\newblock (2007), arXiv:0706.0017 [hep-ph].
%%CITATION = ARXIV:0706.0017;%%

\bibitem{Kramer:2005hw}
M.~Kr{\"a}mer, S.~Mrenna, and D.~E. Soper,
\newblock Phys. Rev. {\bf D73}, 014022 (2006), hep-ph/0509127.
%%CITATION = HEP-PH/0509127;%%

\bibitem{Kurihara:2002ne}
Y.~Kurihara {\em et~al.},
\newblock Nucl. Phys. {\bf B654}, 301 (2003), hep-ph/0212216.
%%CITATION = HEP-PH/0212216;%%

\bibitem{Odaka:2007gu}
S.~Odaka and Y.~Kurihara,
\newblock (2007), hep-ph/0702138.
%%CITATION = HEP-PH/0702138;%%

\bibitem{Giele:2007di}
W.~T. Giele, D.~A. Kosower, and P.~Z. Skands,
\newblock (2007), arXiv:0707.3652 [hep-ph].
%%CITATION = ARXIV:0707.3652;%%

\bibitem{Frixione:2007nu}
S.~Frixione, P.~Nason, and G.~Ridolfi,
\newblock (2007), arXiv:0707.3081 [hep-ph].
%%CITATION = ARXIV:0707.3081;%%

\bibitem{Frixione:2007nw}
S.~Frixione, P.~Nason, and G.~Ridolfi,
\newblock (2007), arXiv:0707.3088 [hep-ph].
%%CITATION = ARXIV:0707.3088;%%

\bibitem{Latunde-Dada:2006gx}
O.~Latunde-Dada, S.~Gieseke, and B.~Webber,
\newblock JHEP {\bf 02}, 051 (2007), hep-ph/0612281.
%%CITATION = HEP-PH/0612281;%%

\bibitem{LatundeDada:2007jg}
O.~Latunde-Dada,
\newblock (2007), arXiv:0708.4390 [hep-ph].
%%CITATION = ARXIV:0708.4390;%%

\bibitem{Sjostrand:2004ef}
T.~Sj{\"o}strand and P.~Z. Skands,
\newblock Eur. Phys. J. {\bf C39}, 129 (2005), hep-ph/0408302.
%%CITATION = HEP-PH/0408302;%%

\bibitem{Gieseke:2003rz}
S.~Gieseke, P.~Stephens, and B.~Webber,
\newblock JHEP {\bf 12}, 045 (2003), hep-ph/0310083.
%%CITATION = HEP-PH 0310083;%%

\bibitem{Lonnblad:2001iq}
L.~L{\"o}nnblad,
\newblock JHEP {\bf 05}, 046 (2002), hep-ph/0112284.
%%CITATION = HEP-PH/0112284;%%

\bibitem{Krauss:2001iv}
F.~Krauss, R.~Kuhn, and G.~Soff,
\newblock JHEP {\bf 02}, 044 (2002), hep-ph/0109036.
%%CITATION = HEP-PH 0109036;%%

\bibitem{Krauss:2005re}
F.~Krauss, A.~Sch{\"a}licke, and G.~Soff,
\newblock Comput. Phys. Commun. {\bf 174}, 876 (2006), hep-ph/0503087.
%%CITATION = HEP-PH/0503087;%%

\bibitem{Gieseke:2004tc}
S.~Gieseke,
\newblock JHEP {\bf 01}, 058 (2005), hep-ph/0412342.
%%CITATION = HEP-PH/0412342;%%

\bibitem{Stephens:2007ah}
P.~Stephens and A.~van Hameren,
\newblock (2007), hep-ph/0703240.
%%CITATION = HEP-PH/0703240;%%

\bibitem{Bauer:2007ad}
C.~W. Bauer and F.~J. Tackmann,
\newblock (2007), arXiv:0705.1719 [hep-ph].
%%CITATION = ARXIV:0705.1719;%%

\bibitem{Bauer:2006mk}
C.~W. Bauer and M.~D. Schwartz,
\newblock (2006), hep-ph/0607296.
%%CITATION = HEP-PH/0607296;%%

\bibitem{Frixione:2003ei}
S.~Frixione, P.~Nason, and B.~R. Webber,
\newblock JHEP {\bf 08}, 007 (2003), hep-ph/0305252.
%%CITATION = HEP-PH/0305252;%%

\bibitem{Frixione:2005vw}
S.~Frixione, E.~Laenen, P.~Motylinski, and B.~R. Webber,
\newblock JHEP {\bf 03}, 092 (2006), hep-ph/0512250.
%%CITATION = HEP-PH/0512250;%%

\bibitem{Frixione:2006gn}
S.~Frixione and B.~R. Webber,
\newblock (2006), hep-ph/0612272.
%%CITATION = HEP-PH/0612272;%%

\bibitem{Frixione:2007zp}
S.~Frixione, E.~Laenen, P.~Motylinski, and B.~R. Webber,
\newblock JHEP {\bf 04}, 081 (2007), hep-ph/0702198.
%%CITATION = HEP-PH/0702198;%%

\bibitem{Catani:1997vz}
S.~Catani and M.~H. Seymour,
\newblock Nucl. Phys. {\bf B485}, 291 (1997), hep-ph/9605323.
%%CITATION = NUPHA,B485,291;%%

\bibitem{Catani:1997vzerr}
S.~Catani and M.~H. Seymour,
\newblock Nucl. Phys. {\bf B510}, 503 (1997),
\newblock Erratum.

\bibitem{Dittmaier:1999mb}
S.~Dittmaier,
\newblock Nucl. Phys. {\bf B565}, 69 (2000), hep-ph/9904440.
%%CITATION = NUPHA,B565,69;%%

\bibitem{Phaf:2001gc}
L.~Phaf and S.~Weinzierl,
\newblock JHEP {\bf 04}, 006 (2001), hep-ph/0102207.
%%CITATION = JHEPA,0104,006;%%

\bibitem{Catani:2002hc}
S.~Catani, S.~Dittmaier, M.~H. Seymour, and Z.~Trocsanyi,
\newblock Nucl. Phys. {\bf B627}, 189 (2002), hep-ph/0201036.
%%CITATION = HEP-PH 0201036;%%

\bibitem{Dokshitzer:1995qm}
Y.~L. Dokshitzer, G.~Marchesini, and B.~R. Webber,
\newblock Nucl. Phys. {\bf B469}, 93 (1996), hep-ph/9512336.
%%CITATION = HEP-PH/9512336;%%

\bibitem{Schumann}
S.~Schumann and F.~Krauss,
\newblock (2007), arXiv:0709.1027 [hep-ph].

\bibitem{Cvitanovic:1980bu}
P.~Cvitanovic, P.~G. Lauwers, and P.~N. Scharbach,
\newblock Nucl. Phys. {\bf B186}, 165 (1981).
%%CITATION = NUPHA,B186,165;%%

\bibitem{Berends:1987cv}
F.~A. Berends and W.~Giele,
\newblock Nucl. Phys. {\bf B294}, 700 (1987).
%%CITATION = NUPHA,B294,700;%%

\bibitem{Mangano:1987xk}
M.~L. Mangano, S.~J. Parke, and Z.~Xu,
\newblock Nucl. Phys. {\bf B298}, 653 (1988).
%%CITATION = NUPHA,B298,653;%%

\bibitem{Kosower:1987ic}
D.~Kosower, B.-H. Lee, and V.~P. Nair,
\newblock Phys. Lett. {\bf B201}, 85 (1988).
%%CITATION = PHLTA,B201,85;%%

\bibitem{Bern:1990ux}
Z.~Bern and D.~A. Kosower,
\newblock Nucl. Phys. {\bf B362}, 389 (1991).
%%CITATION = NUPHA,B362,389;%%

\bibitem{DelDuca:1999rs}
V.~Del~Duca, L.~J. Dixon, and F.~Maltoni,
\newblock Nucl. Phys. {\bf B571}, 51 (2000), hep-ph/9910563.
%%CITATION = HEP-PH 9910563;%%

\bibitem{Maltoni:2002mq}
F.~Maltoni, K.~Paul, T.~Stelzer, and S.~Willenbrock,
\newblock Phys. Rev. {\bf D67}, 014026 (2003), hep-ph/0209271.
%%CITATION = HEP-PH 0209271;%%

\bibitem{Weinzierl:2005dd}
S.~Weinzierl,
\newblock Eur. Phys. J. {\bf C45}, 745 (2006), hep-ph/0510157.
%%CITATION = HEP-PH 0510157;%%

\bibitem{Seymour:1995df}
M.~H. Seymour,
\newblock Comp. Phys. Commun. {\bf 90}, 95 (1995), hep-ph/9410414.
%%CITATION = HEP-PH 9410414;%%

\bibitem{Stirling:1991ds}
W.~J. Stirling,
\newblock J. Phys. {\bf G17}, 1567 (1991).
%%CITATION = JPHGB,G17,1567;%%

\bibitem{Nachtmann:1982xr}
O.~Nachtmann and A.~Reiter,
\newblock Z. Phys. {\bf C16}, 45 (1982).
%%CITATION = ZEPYA,C16,45;%%

\bibitem{Korner:1980pv}
J.~G. K{\"o}rner, G.~Schierholz, and J.~Willrodt,
\newblock Nucl. Phys. {\bf B185}, 365 (1981).
%%CITATION = NUPHA,B185,365;%%

\bibitem{Bengtsson:1988qg}
M.~Bengtsson and P.~M. Zerwas,
\newblock Phys. Lett. {\bf B208}, 306 (1988).
%%CITATION = PHLTA,B208,306;%%

\bibitem{Abreu:1990ce}
DELPHI, P.~Abreu {\em et~al.},
\newblock Phys. Lett. {\bf B255}, 466 (1991).
%%CITATION = PHLTA,B255,466;%%

\bibitem{Pumplin:2002vw}
J.~Pumplin {\em et~al.},
\newblock JHEP {\bf 07}, 012 (2002), hep-ph/0201195.
%%CITATION = HEP-PH/0201195;%%

\bibitem{Stump:2003yu}
D.~Stump {\em et~al.},
\newblock JHEP {\bf 10}, 046 (2003), hep-ph/0303013.
%%CITATION = HEP-PH/0303013;%%

\bibitem{Weinzierl:1999yf}
S.~Weinzierl and D.~A. Kosower,
\newblock Phys. Rev. {\bf D60}, 054028 (1999), hep-ph/9901277.
%%CITATION = PHRVA,D60,054028;%%

\end{thebibliography}
\bibliographystyle{/home/stefanw/latex-style/h-physrev3}

\end{document}